\documentclass[11pt,a4paper]{article}

\usepackage{amsmath,amsfonts}
\usepackage{hyperref,makecell}
\usepackage{graphicx,tikz}
\usetikzlibrary{shapes}
\usetikzlibrary{decorations.pathmorphing}
\usetikzlibrary{decorations.markings}

\def\be{\begin{equation}}
\def\ee{\end{equation}}
\def\bea{\begin{eqnarray}}
\def\eea{\end{eqnarray}}

\newcommand{\SL}{\mathrm{SL}}

\newcommand{\Rset}{\mathbb{R}}
\newcommand{\Ncal}{\mathcal{N}}
\newcommand{\Scal}{\mathcal{S}}

\newcommand{\Vcal}{\mathcal{V}}
\newcommand{\AdS}{\mathrm{AdS}}
\newcommand{\Srm}{\mathrm{S}}
\newcommand{\Cset}{{\,\,{{{^{_{\pmb{\mid}}}}\kern-.47em{\mathrm C}}}}}
\newcommand{\diff}{\mathrm{d}}
\newcommand{\p}{\partial}
\newcommand{\half}{\frac12}
\newcommand{\ra}{\rightarrow}

\newcommand{\beq}{\overset{b}{=}}
\newcommand{\pt}{\tilde{\p}}

\begin{document}

\numberwithin{equation}{section}

\mbox{}

\begin{center}

 {\Large \bf Non-integrability of open strings via Pohlmeyer reduction}\\
\vspace{43pt}

{\large {\mbox{{\bf Divania Carelse$\,{}^{a,b}$}, {\bf Laila Tribelhorn$\,{}^{a,b}$}  \hspace{.1cm} and \hspace{.1cm} {\bf Konstantinos Zoubos$\,{}^{a,c}$}}}}%

\vspace{.5cm}

${}^a$Department of Physics, University of Pretoria\\
Private Bag X20, Hatfield 0028, South Africa

\mbox{}

${}^b$  DSTI-NRF Centre of Excellence in Mathematical and Statistical Sciences (CoE-MaSS)\\South Africa

\mbox{}

${}^c$National Institute for Theoretical and Computational Sciences (NITheCS) \\
Gauteng, South Africa 

 \mbox{}
 
 Email: divania.carelse@tuks.co.za , laila.tribelhorn@tuks.co.za, kzoubos@up.ac.za
\end{center}

\vspace{40pt}

\begin{abstract}
We study integrable and non-integrable boundary conditions for classical strings from the point of view of Pohlmeyer reduction. We first consider the integrable case of strings ending on $Y=0$ non-maximal giant gravitons. Pohlmeyer reduction of such strings leads to a sine-Gordon model, for which a criterion for boundary integrability was provided by Ghoshal and Zamolodchikov. We confirm that the Pohlmeyer field satisfies this condition. We then turn to the main focus of our work, which is open strings on $\AdS_3$. By evaluating the Ghoshal-Zamolodchikov criterion for the corresponding Pohlmeyer-reduced models, we are able to distinguish between integrable and non-integrable cases, in line with expectations from recent analytic and numerical studies. We conclude that Pohlmeyer reduction can provide a useful  perspective on open-string non-integrability. 
\end{abstract}

\vspace{.3cm}

\section{Introduction}

The AdS/CFT correspondence \cite{Maldacena:1997re} between $\Ncal=4$ supersymmetric Yang-Mills theory and IIB string theory on the $\AdS_5\times \Srm^5$ background becomes particularly powerful in the planar limit where the additional feature of integrability arises, see e.g. \cite{Beisert:2010jr} for a review. One can additionally consider the limit of large charges \cite{Gubser:2002tv}, which, on the gravity side, reduces to the study of classical string solutions on the $\AdS_5\times \Srm^5$ geometry. Studies of such classical strings (and their fluctuations) have provided important input into the holographic dictionary, see \cite{Tseytlin:2003ii,Plefka:2005bk,Arutyunov:2009ga,Tseytlin:2010jv} for reviews. 

Given the high amount of symmetry of the background, it is possible and often convenient to map the string sigma model to a simpler scalar model. This process is known as \emph{Pohlmeyer reduction} \cite{Pohlmeyer:1975nb}. For the simplest case where the string motion is restricted to an $\Rset\times \Srm^2$ subspace, where $\Rset$ is the time direction in $\AdS_5$ and $\Srm^2$ is within the five-sphere, Pohlmeyer reduction results in the sine-Gordon model. As the prototypical relativistic 1+1d integrable system, the sine-Gordon model admits a large number of soliton solutions. These solutions can then be classically mapped to the corresponding string configurations. 

It is clearly of interest to ask what happens to integrability when one reduces the amount of symmetry present in the $\AdS_5\times \Srm^5$ setting. This can be done by deforming the geometry itself, by considering an orbifolded geometry, or by introducing boundaries, see \cite{Zoubos:2010kh} for a review. In this work we will focus on the latter case, which is relevant to situations where D-branes are introduced into the background, therefore allowing the existence of an open string sector. The question of integrable boundaries for open strings has been extensively studied, see \cite{Mann:2006rh, Dekel:2011ja} for earlier work and \cite{Linardopoulos:2025ypq} for a recent review focusing on the case of defect CFT. Typically, the starting point is the boundary Yang-Baxter equation \cite{Sklyanin:1987bi,Sklyanin:1988yz}. In the case of the Principal Chiral Model (see e.g. \cite{Driezen:2021cpd} for a pedagogical introduction), it can be expressed in terms of linear relations between the components of the boundary restrictions of the currents \cite{MacKay:2001bh, Delius:2001he,Moriconi:2001xz}. There are also approaches working directly with the Lax connection \cite{Corrigan:1996nt,Aniceto:2017jor,Gombor:2018ppd}, and, in particular, by considering the analytic structure of the Lax connection the authors of \cite{Gil:2026ost} were able to study boundary integrability for more general sigma models, such as cases with mixed fluxes. The extension to the full superstring was recently considered in \cite{Bonomi:2026ygm}. 

The above approaches are concerned with identifying criteria that ensure integrability. It would also be desirable to have a tool to answer the converse question, i.e. to conclusively prove that string motion on a given background, or under certain boundary conditions, is non-integrable. This question is typically approached numerically, by looking for signatures of chaos in the string motion. However, in the closed-string case, there also exists an analytic approach to non-integrability \cite{Basu:2011fw}. It relies on varying around an integrable subsector and checking whether the resulting variational equation is an ODE allowing Liouvillian solutions. Since there exists a classification of such equations (see \cite{MoralesRuiz99} for an introduction) one can confidently argue that if the variational equation is outside this class, classical integrability is not possible. 

Ideally one would like to have a similar tool for the case of open boundaries. However, since the bulk equations of motion of an open string are the same as for closed strings, it doesn't seem that the methods of \cite{Basu:2011fw} are directly applicable to the open case. On the other hand, for models such as sine-Gordon there are established integrability-preserving boundary conditions \cite{Ghoshal:1993tm}, which can provide an analogous criterion. Similarly to the approach of \cite{Basu:2011fw}, one can then argue that if the boundary conditions for the Pohlmeyer-reduced system are not integrable, then those of the full system are not integrable either. 

The idea of studying open strings via their Pohlmeyer reduction is not new, and in particular the case of strings ending on giant gravitons was considered in \cite{Hofman:2007xp} and more extensively addressed in \cite{Ciavarella:2010tp}. The Pohlmeyer reduction of strings ending on various D5-brane configurations was considered in \cite{Correa:2013em}. In this work we take a small further step in developing these techniques, by considering classical strings on $\AdS_3$, where analytic proofs for which types of boundary conditions are integrable, and numerical evidence for which types are not, was recently presented in \cite{Ishii:2023ucw,Ishii:2023qqd}.  

We start in section \ref{Section:GZ} by reviewing the integrable boundary conditions for the sine-Gordon theory and its generalisations. In section \ref{Section:GM} we consider the case of open strings on non-maximal $Y=0$ gravitons. This case is known to be integrable, which is confirmed in our approach as we find integrability-preserving boundary conditions for the Pohlmeyer field. The main part of our study is in Section \ref{Section:AdS3}, where, for the case of strings on $\AdS_3$ with various prescribed boundary conditions, we show agreement between Pohlmeyer reduction and the results of \cite{Ishii:2023ucw,Ishii:2023qqd}. We conclude with several open directions for future work.

\section{Boundary conditions for sine-Gordon-type models} \label{Section:GZ}

Let us start by considering the sine-Gordon model, which is an important example of a relativistic 1+1-dimensional integrable system. For background on the sine-Gordon model and its integrability, we refer to textbooks such as \cite{Babelon:2003qtg}. Its action on the full line is given by
\be
\Scal=\int\diff\tau\int_{-\infty}^\infty\diff\sigma\left(\half \p_\alpha\phi\p^\alpha\phi-\frac{m^2}{\beta^2}\cos(\beta\phi)\right)\;.
\ee
We will set $m=\beta=1$ in the following. We use the worldsheet metric $\eta_{\alpha\beta}=\mathrm{diag}(-1,1)$. Integrability implies the existence of an infinite number of conserved quantities on the full line:
\be\begin{split}
Q_2&=\int\diff\tau\diff\sigma\left(\half (\p_\tau\phi)^2+\half(\p_\sigma\phi)^2+(1-\cos(\phi))\right)\;,\\
Q_4^\pm&=\int\diff\tau\diff\sigma\left(\frac14\left((\p_\tau\pm\p_\sigma)\phi\right)^4-\left((\p_\tau\pm\p_\sigma)^2\phi\right)^2-((\p_\tau\pm\p_\sigma) \phi)^2\cos\phi\right)\;,\\
&\;\vdots
\end{split}
\ee
with $Q_2$ being the Hamiltonian and $Q_4$ the first higher charge, which can be constructed by integrability-based methods \cite{Babelon:2003qtg}. Ghoshal and Zamolodchikov \cite{Ghoshal:1993tm} studied the conservation of the higher charges of the sine-Gordon model on the half-line, depending on the boundary conditions imposed at $\sigma=0$. Assuming that the boundary action does not have explicit time dependence, one writes:
\be
\Scal=\int\diff\tau\int_{-\infty}^0\diff\sigma\left(\half \p_\alpha\phi(\tau,\sigma)\p^\alpha\phi(\tau,\sigma)-\cos\phi(\tau,\sigma)\right)+\delta(\sigma)\int\diff\tau \Vcal(\phi(\tau,\sigma))\;,\ee\
which leads to the boundary condition:
\be
\p_\sigma \phi\beq-\frac{d\Vcal(\phi)}{d\phi}\;,
\ee
where $\beq$ indicates an expression valid on the boundary $\sigma=\sigma_b$. One then looks for a condition on $\Vcal(\phi)$ such that the boundary restriction of the integrand in $Q_4^++Q_4^-$ is a total $\tau$ derivative, leading to conservation of a higher charge on the half-line. The resulting condition obtained in \cite{Ghoshal:1993tm} is
\be \label{GZ}
\p_\sigma \phi\beq-4K \sin\left(\frac{\phi-\phi_0}{2}\right)\;,
\ee
with $K,\phi_0$ constants and $\phi$ taking its boundary value, $\phi=\phi(\tau,0)$. This condition interpolates in an integrable way between the pure Neumann boundary condition at $K=0$ and the pure Dirichlet condition as $K\ra\infty$. 

Even though the original derivation in \cite{Ghoshal:1993tm} was for the first higher charge, by applying the boundary integrability formalism of \cite{Sklyanin:1987bi} it can be shown \cite{MacIntyre:1994sr} that  (\ref{GZ}) is the most general sine-Gordon boundary condition compatible with integrability.\footnote{This statement assumes no additional degrees of freedom at the boundary. See e.g. \cite{Baseilhac:2001zd} for an integrable case where sine-Gordon is coupled to a mechanical system at the boundary.} Solutions of the limiting (Neumann or Dirichlet) boundary conditions can be mapped to those of the more general (\ref{GZ}) by a sequence of B\"acklund transformations \cite{Saleur:1994yh}.

The integrable boundary condition (\ref{GZ}) implies that solitons approaching the boundary are reflected as (anti-)solitons, with a possible phase shift. Conversely, when imposing boundary conditions outside this class, one expects boundary scattering to lead to chaotic behaviour. In the sine-Gordon model, a detailed numerical study of the case of (non-integrable) Robin boundary conditions was performed in  \cite{Arthur:2015mva} and highlighted the sensitivity of the reflected data on the incoming data.

The Ghoshal-Zamolodchikov boundary condition can be extended beyond sine-Gordon theory to the wider class of affine Toda theories \cite{Corrigan:1994ft,Corrigan:1994np,Corrigan:1995wx,Bowcock:1995vp} and in particular also to the sinh-Gordon and cosh-Gordon cases which, as we will see, are relevant for strings on $\AdS_3$. It turns out that in the higher Toda theories boundary integrability imposes additional constraints on the coefficients, however for sinh/cosh-Gordon the analytic continuation of (\ref{GZ}) is still the relevant criterion. So in the following we will take the boundary condition
\be \label{GZsinhcosh}
\p_\sigma \phi\beq 2K \left(e^{\frac{\phi}2}\pm e^{-\frac{\phi}2}\right)\;,
\ee
with $\pm$ for cosh/sinh-Gordon respectively, as the benchmark for boundary integrability in these models. We will still refer to (\ref{GZsinhcosh}) as Ghoshal-Zamolodchikov boundary conditions. 

We note that for integrability the conditions (\ref{GZ}) or (\ref{GZsinhcosh}) need to be satisfied at each boundary. Therefore, in the following, we will only focus on one of the two boundaries of our string solutions. 

\section{Strings on giant gravitons} \label{Section:GM}

As a warm-up, in this section we consider open strings ending on giant gravitons. This is a well-known scenario for open boundary conditions to appear in the $\AdS_5\times \Srm^5$ setting. The material in this section is not new but we will present it in a way that informs the study of $\AdS_3$ in the following section. 

Giant gravitons \cite{McGreevy:2000cw,Grisaru:2000zn} are D3-branes wrapped on cycles in a 10d geometry, stabilised by their angular momentum. We will consider the specific case of the branes wrapping the time direction within $\AdS_5$ and an $\Srm^3 \subset \Srm^5$. In the gauge theory, they have an interpretation as determinant or subdeterminant operators made up of one of the complex scalar fields of $\Ncal=4$ SYM \cite{Balasubramanian:2001nh}. Open-string configurations ending on giant gravitons have been widely studied, see \cite{Berenstein:2006qk,Mann:2006rh,Hofman:2007xp,Dekel:2011ja} for some foundational works.  For additional references on open strings in the AdS/CFT context, we refer to the review \cite{Zoubos:2010kh}.

To understand how the D3-branes are embedded in the five-sphere, consider the homogeneous coordinates of the (unit) $\Srm^5$, satisfying $\sum_i(X^i)^2=1$ with $i=1,\ldots, 6$. They can be combined into complex coordinates
\be
X=X^1+iX^2\;,\;Y=X^3+i X^4 \;,\;Z=X^5+i X^6\;.
\ee
Following \cite{Hofman:2007xp}, we can label the different choices of $\Srm^3$ inside $\Srm^5$ wrapped by the brane as 
\be\begin{split}
X=0&\;\; \text{graviton}: |Y|^2+|Z|^2=r^2\;, \\
Y=0&\;\; \text{graviton}: |X|^2+|Z|^2=r^2\;, \\
Z=0&\;\; \text{graviton}: |X|^2+|Y|^2=r^2\;. \\
\end{split}
\ee
The maximal case is when $r=1$. These are of course equivalent embeddings, however as mentioned stability of the D-branes also requires them to have angular momentum on the $S^5$. Choosing the rotation to be in the $X^5-X^6$ (i.e. $Z$) plane, the $Z=0$ graviton is distinguished from the others in that it is pointlike along its direction of motion. It is conventional to choose the $Y=0$ graviton to represent the alternative configuration, with the $X=0$ case being related by a rotation. 
These two choices ($Y=0$ and $Z=0$) lead to different boundary conditions once we introduce classical strings whose motion is restricted to a specific submanifold of $\Srm^5$. We will be interested in classical string solutions on $\Rset\times \Srm^2$, with $\Rset$ the time direction within $\AdS_5$ and $\Srm^2\subset \Srm^5$. Let us choose the $\Srm^2$ to be the one parametrised by $X^3,X^5,X^6$. Then the sphere on which the string moves is given by
\be \label{S2emb}
(X^3)^2+(X^5)^2+(X^6)^2=1\;.
\ee
Note that in this case the string worldsheet extends in the $Y$ direction and is therefore transverse to the worldvolume of a $Y=0$ graviton. An alternative configuration would be for the $\Srm^2$ to contain $X^1$ or $X^2$ instead of $X^3$, placing the worldsheet within the $Y=0$ graviton's worldvolume \cite{Hofman:2007xp,Ciavarella:2010tp}. Here we will only consider the transverse case (\ref{S2emb}). 

An important class of classical solutions on $\Rset\times\Srm^2$ are the \emph{giant magnons} \cite{Hofman:2006xt}, which in the above notation rotate in the $X^5-X^6$ plane with their endpoints on the equator at $X^3=0$.\footnote{In the original formulation \cite{Hofman:2006xt}, giant magnons were thought of as halves of closed strings, but can also be usefully interpreted as open strings \cite{Hofman:2007xp}. This is in line with their gauge theory interpretation as open chains ending on (sub)determinant operators.}  The angular separation of the endpoints is related to the momentum of the magnon in the dual spin-chain picture. We see that the original giant magnon is an open string with endpoints on a maximal $Y=0$ giant graviton. Since in that case $|Y|=0$, the graviton meets the $\Srm^2$ at $X^3=0$, i.e. the equator. 

As we will review in the next section, open strings with maximal $Y=0$ boundary conditions are integrable. This is in agreement with 
the integrability of spin chains with $Y=0$ boundary conditions on the gauge theory side, first shown in \cite{Berenstein:2005vf}, with the Bethe ansatz equations constructed in \cite{Galleas:2009ye, Nepomechie:2009zi}. Further work on the spin chain side \cite{MacKay:2010ey} showed that the $Y=0$ boundary condition preserves a subset of the Yangian algebra. The Y-system \cite{Bajnok:2012xc} and spectral curve \cite{Bajnok:2013sza} for this case have also been considered. Agreement of correlation functions on both sides of the correspondence was shown in \cite{Bak:2011yy}.

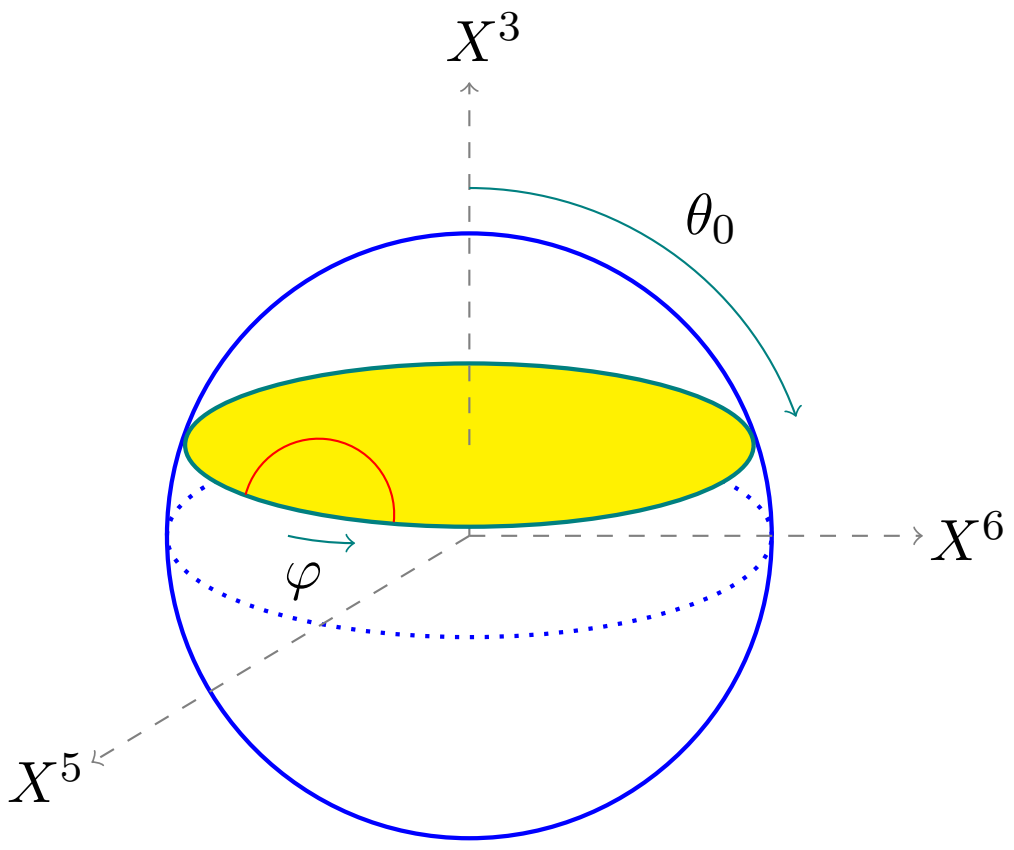
\begin{figure}[ht]
\begin{center}\begin{tikzpicture}
  \draw[-,thick,blue] (2,0) arc (0:360:2cm);
\draw[-,thick,blue,dotted] (2,0) arc (0:360:2cm and 0.67cm);
\draw[->,dashed,gray] (0,0)--(3,0);
\draw[->,dashed,gray] (0,0)--(-2.5,-1.5);
\draw[->,dashed,gray] (0,0)--(0,1);
\filldraw[yellow] (1.88,0.6) arc (0:360:1.88cm and 0.54cm);
\draw[-,thick,teal] (1.88,0.6) arc (0:360:1.88cm and 0.54cm);
\draw[->,dashed,gray] (0,0.6)--(0,3);
\draw[-,red] (-0.5,0.09) arc (-6:165:0.5cm and 0.5cm);
\draw[->,teal] (0,2.3) arc (90:20:2.3cm);
\draw[->,teal] (-1.2,0) arc (250:270:1.3cm and 0.8cm);
\node at (-2.8,-1.6) {$X^5$};
\node at (3.3,0) {$X^6$};
\node at (0.1,3.3) {$X^3$};
\node at (1.6,2.1) {$\theta_0$};\node at (-1.1,-0.3) {$\varphi$};
\end{tikzpicture}\end{center}
\caption{A sketch of the non-maximal $Y=0$ graviton (yellow plane) intersecting with the $\Srm^2$ on which the string motion is restricted. A string solution (red, schematic) satisfies the Dirichlet condition $\theta(\tau,\sigma_b^{(L,R)})=\theta_0$, where $\sigma_b^{(L,R)}$ indicates the left and right endpoints, and a Neumann condition in the $\varphi$ direction. The maximal case is $\theta_0=\pi/2$ where the graviton is located at the equator of the $\Srm^2$, while the ``maximally non-maximal'' case is at $\theta_0=0$, where the graviton is located at the pole.} \label{Fig:nmY0}
\end{figure}

One can move away from the maximal $Y=0$ case by taking $r<1$ in the $Y=0$ case above. Since $|X|^2+|Y|^2+|Z|^2=1$ but $|X|^2+|Z|^2=r^2$, we have that $|Y|^2=1-r^2$. Then the intersection of the graviton with the $\Srm^2$ happens at a fixed latitude $X^3=\pm \sqrt{1-r^2}$. A sketch of the configuration one obtains is shown in Fig. \ref{Fig:nmY0}. This configuration was argued in \cite{Berenstein:2006qk} to be integrable, both from the classical string and dual spin chain perspectives. A detailed analysis and construction of open string solutions in the non-maximal $Y=0$ case was performed in \cite{Ciavarella:2010tp}.

Although we will only focus on $Y=0$ gravitons in this work, it is worth mentioning that the $Z=0$ case is rather different. The maximal $Z=0$ graviton is pointlike in the $\Srm^2$, intersecting only at the poles $X^3=\pm 1$. Therefore a giant magnon (with an endpoint on the equator) needs to be connected to the pole via what is known as a boundary giant magnon. Such solutions were constructed in \cite{Ciavarella:2010je}. On the spin chain side, these degrees of freedom arise as boundary impurities where the chain connects to the determinant operator. Scattering on the corresponding spin chain has been studied in \cite{Ahn:2007bq,Bak:2008xq}, and three-point functions in \cite{Bak:2011yy}. Scattering with \emph{non-maximal} $Z=0$ boundary conditions (where the D-brane intersects the $\Srm^2$ away from the poles and hence drags the string endpoints as it rotates)  has been less studied and integrability in this case is not clear \cite{Ciavarella:2010tp}.

\subsection{Pohlmeyer reduction on $\Rset\times \Srm^2$} \label{PohlmeyerRS2}

Let us briefly review the Pohlmeyer reduction of strings on $\Rset\times \Srm^2$  to sine-Gordon theory \cite{Pohlmeyer:1975nb}. As discussed, we will take the string to live in the $\Srm^2$ defined by the Cartesian coordinates $X^i$, with $i=3,5,6$. They can be expressed in terms of intrinsic coordinates as
\be
X^5=\sin\theta\cos\varphi\;\;,\;\;X^6=\sin\theta\sin\varphi\;\;,\;\; X^3=\cos\theta\;.
\ee
The $\Rset$ is the time direction, which we take to depend on the worldsheet coordinates as $t(\tau,\sigma)=\tau$. Then the Polyakov string action in conformal gauge is
\be\begin{split}
\Scal&=-\frac{1}{2\pi\alpha'}\int\diff\tau\diff\sigma\left(-\p_\alpha t\p^\alpha t+ \delta_{ij}\p^\alpha X^i\p_\alpha X^j\right)\\
&=-\frac{1}{2\pi\alpha'}\int\diff\tau\diff\sigma\left(1+\p_\alpha \theta\p^\alpha\theta+\sin^2\theta \p_\alpha \varphi\p^\alpha\varphi\right)\;. 
\end{split}
\ee
The string equations of motion are
\be
\p_{\tau\tau}X^i -\p_{\sigma\sigma}X^i+\cos\phi X^i=0\;,
\ee
where we have defined the Pohlmeyer field $\phi$ as
\be \label{S2Pohlmeyer}
\cos\phi=\p_\tau X^i\p_\tau X^i-\p_\sigma X^i\p_\sigma X^i\;.
\ee
The preservation of the conformal gauge condition requires us to also impose the Virasoro constraints
\be
(\p_\tau X^i)(\p_\tau X^i)+(\p_\sigma X^i)(\p_\sigma X^i)=1 \;\;\text{and}\;\;\p_\tau X^i\p_\sigma X^i=0\;,
\ee
with the last condition implying that the motion of the string endpoint is perpendicular to the string itself. Combining (\ref{S2Pohlmeyer})
with the Virasoro constraints we find the conditions
\be \label{sincos}
\sin^2\frac\phi2=(\p_\sigma X^i)^2\;\;\text{and}\;\; \cos^2\frac\phi2=(\p_\tau X^i)^2\;.
\ee
The equations of motion together with the Virasoro constraints lead to the sine-Gordon equation for the Pohlmeyer field:
\be
\p_{\tau\tau}\phi-\p_{\sigma\sigma}\phi+\sin\phi=0\;.
\ee
At the level of solutions, one can therefore map classical strings on $\Rset\times \Srm^2$ to sine-Gordon solitons. For instance, the giant magnon solution of \cite{Hofman:2006xt} reduces to a sine-Gordon kink. See \cite{Klose:2008rx} for a large range of examples both on the infinite line and on the finite interval. The inverse procedure of uplifting sine-Gordon solitons to solutions of the sigma model is more involved but was also addressed in \cite{Klose:2008rx}, as well as (with a different approach) in \cite{Katsinis:2018zxi,Katsinis:2018ewd,Katsinis:2019oox}. 

The Pohlmeyer reduction can be generalised to $\Srm^3$, leading to complex sine-Gordon \cite{Chen:2006gea, Okamura:2006zv}, as well as to more general symmetric target spaces, see \cite{Grigoriev:2007bu,Miramontes:2008wt,Hollowood:2009tw,Hollowood:2010dt,Katsinis:2022hyn}, with the analogue of sine-Gordon in these cases being the symmetric-space sine-Gordon theory \cite{Bakas:1995bm}. 

In the following section we will turn to the Pohlmeyer reduction in the open-string case.

\subsection{Pohlmeyer reduction for non-maximal $Y=0$ gravitons}

Pohlmeyer reduction for open strings ending on maximal $Y=0$ giant gravitons was considered already in \cite{Hofman:2007xp}. As shown there, it corresponds to $\p_\sigma\phi\beq0$, i.e. a pure Neumann boundary condition for the Pohlmeyer field. The non-maximal $Y=0$ case was considered in \cite{Ciavarella:2010tp}, with a view to the construction of explicit open string solutions with this boundary condition, although in practice the method used to do so was not directly uplifting sine-Gordon solitons, but rather applying the method of images to a (closed-string) 3-magnon solution previously constructed in \cite{Kalousios:2008gz}. Here we are not interested in solutions, but only in confirming that the Pohlmeyer boundary condition for the generic non-maximal $Y=0$ case takes the Ghoshal-Zamolodchikov form (\ref{GZ}), as would be expected from the integrability of this boundary configuration. Our presentation below closely follows that of \cite{Ciavarella:2010tp}. 

As mentioned (see Fig. \ref{Fig:nmY0}), the non-maximal $Y=0$ graviton intersects the $\Srm^2$ of the string on a circle of fixed latitude, $\theta(\tau,\sigma_b)= \theta_0$ with the string endpoints free to move along the azimuthal angle $\varphi$. So we have a Dirichlet condition for $\theta$ and a Neumann condition for $\varphi$:\footnote{As mentioned, we only consider the $Y=0$ case where the string worldsheet is transverse to the brane worldvolume.}
\be
\p_\tau \theta\beq 0\;\;,\;\; \p_\sigma\varphi\beq 0\;.
\ee
From (\ref{sincos}) we find that (by choice of sign) the Pohlmeyer field $\phi$ on the boundary is simply
\be
\sin\frac{\phi}2\beq\p_\sigma \theta\;.
\ee
We note that $\p_\sigma\theta$ is generically $\tau$-dependent, so this is not a Dirichlet condition on $\phi$. Therefore we need to evaluate the $\sigma$ derivative of $\phi$, by first taking the derivative of its bulk expression and then restricting to the boundary:
\be \label{dsigmaYzero1}
\sin\phi ~\p_\sigma \phi\beq4\p_\sigma \theta \p_{\sigma\sigma}\theta\;.
\ee
We can obtain an expression for $\p_{\sigma\sigma}\theta$ by restricting the equations of motion to the boundary, which gives
\be
\p_{\sigma\sigma}\theta\beq -\sin\theta\cos\theta_0 (\p_\tau\varphi)^2\;.
\ee
The derivative of the angle $\varphi$ can be read off from the Virasoro constraints:
\be
(\p_\tau\varphi)^2\beq \frac{1}{\sin^2\theta_0} \left(1-(\p_\sigma \theta)^2\right) \beq\frac{1}{\sin^2\theta_0}\left(\cos\frac{\phi}{2}\right)^2\;.
\ee
Substituting in (\ref{dsigmaYzero1}), we find
\be
\p_\sigma\phi\beq -2\cot\theta_0 \cos\left(\frac{\phi}{2}\right)\beq-2\cot\theta_0 \sin\left(\frac{\phi}{2}-\frac\pi2\right)\;.
\ee
This is precisely the Ghoshal-Zamolodchikov integrable boundary condition (\ref{GZ}). As $\theta_0=\frac\pi2$ we recover the Neumann boundary condition corresponding to the maximal $Y=0$ case, while in the ``maximally non-maximal'' case of  $\theta_0=0$ we obtain a Dirichlet boundary condition.  This is in accordance with the discussion in \cite{Ciavarella:2010tp}, where it was noted that this limit of the non-maximal $Y=0$ boundary condition matches that of the maximal $Z=0$ case.\footnote{As can be seen from Fig. \ref{Fig:nmY0}, as $r\ra 0$  the non-maximal $Y=0$ graviton approaches the pole of the $\Srm^2$, which is the location of the maximal $Z=0$ graviton.} As expected, for intermediate values of $\theta_0$, the boundary condition interpolates between these two extremes in an integrable way.

Having confirmed via Pohlmeyer reduction that the non-maximal $Y=0$ magnon boundary condition is integrable, in agreement with \cite{Berenstein:2006qk} and \cite{Ciavarella:2010tp}, we can now more confidently proceed to the analogous analysis for strings on $\AdS_3$.

\section{Open strings on AdS$_3$} \label{Section:AdS3}

Let us now move on to the main focus of this work, which is the (non)integrability of various boundary conditions on $\AdS_3$. This question was studied recently in \cite{Ishii:2023ucw,Ishii:2023qqd}, by applying the formalism for integrable boundary conditions in the Principal Chiral Model initially developed in \cite{MacKay:2001bh,Delius:2001he}. Recall that the PCM on $\AdS_3$, with isometry $\SL(2,\Rset)_L\times \SL(2,\Rset)_R$, has two conserved currents (one with respect to each $\SL(2,\Rset)$ factor). The condition for boundary integrability can be translated to whether the currents satisfy any linear relations, which one can try to verify for specific choices of boundary conditions for each of the $\AdS_3$ coordinates. Let us then start by writing the global $\AdS_3$ metric as
\be\label{AdS3metric}
\diff s^2=-\cosh^2\rho~\diff t^2+\diff \rho^2+\sinh^2\rho~\diff\theta^2\;.
\ee
One can then classify the open-string boundary conditions according to whether each of the coordinates satisfies a Neumann (N) or Dirichlet (D) boundary condition. Among the cases considered in \cite{Ishii:2023ucw}, we will focus on the following cases:\footnote{We do not consider cases with a Dirichlet condition in the time direction, which were also analysed in \cite{Ishii:2023ucw}.}
\be \label{AdSbc}
 (t(\tau,\sigma),\rho(\tau,\sigma),\theta(\tau,\sigma)) : \text{(N,N,N), (N,N,D), (N,D,N), (N,D,D)}
\ee
By analysing the criteria of \cite{MacKay:2001bh,Delius:2001he} applied to the $\AdS_3$ case, it was shown in \cite{Ishii:2023ucw} that the NNN and NND conditions are integrable. On the other hand, in the NDN and NDD cases the currents did not satisfy any linear relations and therefore integrability was inconclusive. However, the question of integrability was also studied numerically by observing the presence of turbulence when perturbing a string solution subject to one of the above boundary conditions. It was shown in \cite{Ishii:2023ucw} for NDD and in \cite{Ishii:2023qqd} for NDN that turbulence is indeed present, providing a strong indication that those boundary conditions break integrability. 

To obtain a different perspective on open boundary conditions on $\AdS_3$, we will consider the same cases as above from the perspective of Pohlmeyer reduction. In particular, we will ask what boundary conditions arise for the Pohlmeyer field under each of these string boundary conditions. So, after reviewing Pohlmeyer reduction for $\AdS_3$ in the following section, we will consider each of the above cases in turn.

\subsection{Pohlmeyer reduction for $\AdS_3$}

The analogue of Pohlmeyer reduction for $\AdS_3$ (as well as $\mathrm{dS}_3$) was first considered in \cite{DeVega:1992xc}, and revisited in the AdS/CFT context in \cite{Hayashi:2007bq, Jevicki:2007aa, Irrgang:2012xb,Bakas:2016jxp}. These works are all in the closed-string case, while to our knowledge the open-string case has not been considered in the literature. As the construction has some important differences to the $\Rset\times \Srm^2$ case discussed in Section \ref{PohlmeyerRS2}, in this section we will briefly review its main features before considering the boundary conditions in the next section. Our notation combines aspects of that of \cite{Jevicki:2007aa} and \cite{Bakas:2016jxp}, to which we also refer for further details and the construction of solutions. 

Let us parametrise $\AdS_3$ by the $\Rset^{(2,2)}$ homogeneous coordinates $Y^\mu$, $\mu=0,1,2,3$. They satisfy
\be \label{AdS3hom}
\eta_{\mu\nu} Y^\mu Y^\nu=-(Y^0)^2-(Y^1)^2+(Y^2)^2+(Y^3)^2=-1\;.
\ee
We can express them in terms of intrinsic coordinates as
\be 
Y^\mu=\left(\begin{array}{c} \cosh\rho\cos t\\ \cosh\rho\sin t \\ \sinh\rho\cos\theta \\ \sinh\rho \sin\theta
\end{array}\right)\;,
\ee
which leads to the metric (\ref{AdS3metric}). 
In the following it will be convenient to use lightcone coordinates
\be
x^\pm=\frac{\sigma\pm\tau}2 \;\;\Rightarrow \;\; \p_\pm=\p_\sigma\pm \p_\tau\;.
\ee
Expressed in these coordinates, the string equations of motion are
\be \label{EomY}
\p_+\p_- Y^\mu=(\p_+Y\cdot \p_- Y)  Y^\mu=e^\alpha  Y^\mu\;,
\ee
where we have defined (with ``$\cdot$'' denoting  contraction with the metric in (\ref{AdS3hom}))
\be \label{eA}
e^\alpha=\p_+Y\cdot \p_- Y\;.
\ee
As usual, we also need to impose the Virasoro constraints
\be
\p_\pm Y\cdot \p_\pm Y=0\;.
\ee
The definition of $e^\alpha$ is essentially the same as that of the Pohlmeyer field in (\ref{S2Pohlmeyer}), however the equation it satisfies can be seen to be 
\be \label{dpmea}
  \p_+\p_- e^\alpha=e^{2\alpha}+\p_+^2 Y \cdot \p_-^2 Y\;.
\ee
This is not immediately expressible in terms of $\alpha$, so in this case the approach of Section \ref{PohlmeyerRS2} will need to be modified slightly. Clearly,  a convenient expression for the vectors $\p_\pm^2 Y^\mu$ would be helpful. To find such an expression, it will be useful to construct a basis in the space of 4-vectors, in which any arbitrary vector can be expressed. Differentiating (\ref{AdS3hom}), we find 
\be
Y\cdot Y=-1 \;\;\ra \;\; \p_\pm Y \cdot Y =0\;,
\ee
so $\p_\pm Y^\mu$ are orthogonal to $Y^\mu$. They are also null by the Virasoro constraints, but not orthogonal to each other, since $\p_+ Y\cdot \p_- Y=e^\alpha$. To have a complete basis, we need one more linearly-independent vector, which needs to be orthogonal to all three. We can define it as
\be
B_\mu=e^{-\alpha} \epsilon_{\mu\nu\kappa\lambda} Y^\nu \p_+Y^\kappa \p_- Y^\lambda\;,
\ee
which satisfies
\be
B_\mu Y^\mu=0\;,\;B_\mu \p_+Y^\mu=0\;,\;B_\mu \p_-Y^\mu=0\;\;,\; B_\mu B^\mu=1\;,
\ee
where the Virasoro constraints are needed to show the last equation. Any vector can now be expressed in the basis $\{Y^\mu,\p_+ Y^\mu,\p_- Y^\mu,B^\mu\}$. To do this for $\p_\pm Y^\mu$ we start with
\be
B\cdot\p_+ Y=0 \; \overset{\p_+}{\ra} \; \p_+B \cdot \p_+Y+B\cdot \p_+^2 Y=0\;.
\ee
We then define
\be
u^+=B\cdot \p_+^2 Y=-\p_+B \cdot \p_+Y\;,
\ee
and similarly 
\be
u^-=B\cdot \p_-^2 Y=-\p_-B\cdot \p_- Y\;.
\ee
It is straightforward to check that the $u^\pm$ satisfy
\be
\p_-u^+=\p_+u^-=0\;,
\ee
i.e. $u^+=u^+(x^+)$ and $u^-=u^-(x^-)$. 

From (\ref{EomY}) we can also find
\be
\p_\pm^2Y \cdot\p_\mp Y=e^{\alpha}\p_\pm\alpha\;,
\ee
and it is easy to check that $\p_+^2 Y\cdot Y=0$ and $\p_+^2 Y \cdot \p_+Y=0$. Then we can express
\be \label{updef}
\p_+^2 Y^\mu=(\p_+\alpha) \p_+ Y^\mu+u^+ B^\mu\;,
\ee
and
\be \label{umdef}
\p_-^2 Y^\mu=(\p_-\alpha) \p_- Y^\mu+u^- B^\mu\;.
\ee
Substituting these expressions in (\ref{dpmea})  we finally obtain
\be \label{ualphaequation}
\p_+\p_-\alpha=e^\alpha+u^+u^- e^{-\alpha}\;.
\ee
At this point we have three possibilities depending on the sign of the product $u^+u^-$. 
If $u^+$ and/or $u^-$ are zero, we obtain the Liouville equation. If $u^+u^-$ is negative, we can rescale the coordinates $x^\pm\ra \tilde{x}^\pm$, such that
\be \label{rescaledderivatives}
\p_+=\sqrt{|u^+|} \tilde{\p}_+\;\;,\;\;\p_-=\sqrt{|u^-|} \tilde{\p}_-\;,
\ee
and define
\be
e^\phi=\tilde{\p}_+ Y\cdot \tilde{\p}_- Y\;,
\ee
which is related to $e^\alpha$ as
\be
e^\phi=\frac{1}{\sqrt{u^+u^-}}\p_+ Y\cdot \p_- Y=\frac{1}{\sqrt{u^+u^-}} e^{\alpha}\;.
\ee
This results in the sinh-Gordon equation for $\phi$, which is the true Pohlmeyer field for $\AdS_3$:
\be \label{sinhG}
\tilde{\p}_+\tilde{\p}_-\phi=e^{\phi}-e^{-\phi}\;.
\ee
Similar manipulations for the case $u^+u^->0$ lead to the cosh-Gordon equation
\be \label{coshG}
\tilde{\p}_+\tilde{\p}_-\phi=e^{\phi}+e^{-\phi}\;.
\ee
All these cases (Liouville, sinh-Gordon and cosh-Gordon) are of course integrable 1+1 dimensional systems, in line with the integrability of closed string theory on $\AdS_3$. The above construction can be generalised to larger sectors, such as $\AdS_3\times \Srm^1$ \cite{Hayashi:2007bq}, where one instead obtains \emph{complex} sinh- or cosh- Gordon. Pohlmeyer reduction for the full $\AdS_5\times \Srm^5$ sigma model has been studied in \cite{Grigoriev:2007bu,Mikhailov:2007xr,Hoare:2009fs,Hoare:2011fj,Hoare:2012nx}.

Of course, as discussed, the introduction of boundaries can break integrability. In the following sections, we will be interested in how the different possible boundary conditions on $t,\rho,\theta$ listed in (\ref{AdSbc}) map to boundary conditions on the Pohlmeyer field $\phi$. As it is only in the rescaled coordinates (\ref{rescaledderivatives}) that we obtain (\ref{sinhG}) or (\ref{coshG}), in order to compare with the GZ condition (\ref{GZsinhcosh}), we need to express the boundary conditions in terms of $\tilde{\sigma}$:
\be \label{AdSPohlmeyerbc}
\pt_{\sigma}e^\phi\beq\half (\pt_++\pt_-)e^\phi\beq\half\left(\frac{1}{\sqrt{|u^+|}}\p_++\frac{1}{\sqrt{|u^-|}}\p_-\right)\frac{1}{\sqrt{u^+u^-}} e^{\alpha}\;.
\ee
We are now ready to examine the four boundary conditions (\ref{AdSbc}) in turn.

\subsection{The NNN case} \label{Sec:NNN}

In this case we allow the endpoints to move freely in all directions of $\AdS_3$, corresponding to Neumann boundary conditions for all the coordinates:
\be \label{bcNNN}
\p_\sigma t(\tau,\sigma)\beq \p_\sigma \rho(\tau,\sigma) \beq \p_\sigma \theta(\tau,\sigma)\beq 0\;.
\ee
As discussed in \cite{Ishii:2023ucw}, it is straightforward to show that these boundary conditions satisfy the PCM criteria of \cite{MacKay:2001bh,Delius:2001he} and therefore the NNN case is integrable. We will now consider these conditions from the viewpoint of the Pohlmeyer-reduced field. 

It is easiest to study this case in Cartesian coordinates. Assuming for simplicity that the boundary is at $\sigma=0$, we can write the Neumann conditions (\ref{bcNNN}) as
\be
 \p_\sigma Y^\mu=0+\sigma \p_{\sigma\sigma}Y^\mu+\cdots\;\;,\;\;\p_{\sigma\tau}Y^\mu=0+\sigma \p_{\sigma\sigma\tau}Y^\mu+\cdots
 \ee
Then the Virasoro constraints can be expanded around the boundary as
 \be
 \p_{\pm} Y\cdot \p_{\pm}Y=(\p_\sigma\pm\p_\tau)Y\cdot (\p_\sigma\pm\p_\tau)Y=(\p_\tau Y)^2\pm2\sigma (\p_\tau Y)\cdot(\p_{\sigma\sigma}Y)+\sigma^2(\p_{\sigma\sigma} Y)^2+\cdots
 \ee
Combining the two constraints, we find that
 \be
 (\p_\tau Y)^2=-\sigma^2(\p_{\sigma\sigma} Y)^2+\cdots
 \ee
Therefore we see that 
 \be
 e^\alpha=(\p_\sigma Y)^2-(\p_\tau Y)^2=\sigma^2 (\p_{\sigma\sigma} Y)^2-(\p_\tau Y)^2=2\sigma^2(\p_{\sigma\sigma} Y)^2+\cdots
 \ee
 So $e^\alpha$ vanishes quadratically at the boundary. 
 To translate the condition on $\alpha$ into a condition on the Pohlmeyer field $\phi$, we need to also consider the boundary behaviour of $u^\pm$. First note that
  \be
 B_\mu=e^{-\alpha}\epsilon_{\mu\nu\kappa\lambda}Y^\nu(\p_\sigma+\p_\tau)Y^\kappa(\p_\sigma-\p_\tau)Y^\lambda=2\sigma e^{-\alpha}\epsilon_{\mu\nu\kappa\lambda}Y^\nu\p_\tau Y^\kappa\p_{\sigma\sigma}Y^\lambda+\cdots
 \ee
This expression diverges generically, due to the double pole in the $e^{-\alpha}$ prefactor. However, substituting into $u^+$ we obtain
 \be
 u^+=B_\mu \p_+^2 Y^\mu=4\sigma^2e^{-\alpha}\epsilon_{\mu\nu\kappa\lambda}Y^\mu\p_\tau Y^\kappa \p_{\sigma\sigma}Y^\lambda \p_{\sigma\sigma\tau}Y^\mu\;,
\ee
where we used the equations of motion
 \be
 \p_{\sigma\sigma}Y^\mu-\p_{\tau\tau}Y^\mu=e^\alpha Y^\mu\;,
 \ee
to express 
 \be
 \p_+^2 Y^\mu=(\p_{\sigma\sigma}+2\p_{\sigma\tau}+\p_{\tau\tau})Y^\mu=(2\p_{\sigma\sigma}-e^{\alpha}+2\sigma\p_{\sigma\sigma\tau}+\cdots)Y^\mu\;,
 \ee
 and used the antisymmetry of the $\epsilon$ tensor. We see that generically (i.e. unless $Y^\mu$ or some of its derivatives vanish or are proportional) $u^+$ is finite at the boundary, and in particular
 \be
 u^+\beq\frac{\epsilon_{\mu\nu\kappa\lambda}Y^\mu\p_\tau Y^\kappa \p_{\sigma\sigma}Y^\lambda \p_{\sigma\sigma\tau}Y^\mu}{(\p_{\sigma\sigma} Y)^2}\;.
 \ee
 Similarly we find that $u^-\beq -u^+$, i.e. the equation (\ref{ualphaequation}) can be mapped to sinh-Gordon, at least close to the boundary. Therefore, given that $e^\alpha\beq 0$, generically we find that
 \be
 e^\phi=\frac{e^\alpha}{\sqrt{|u^+u^-|}} \beq 0\;,
 \ee
i.e. $\phi\ra -\infty$ at the boundary.  We have found a Dirichlet boundary condition, from which we conclude that the NNN case is integrable. 

\paragraph{Example: The GKP solution}

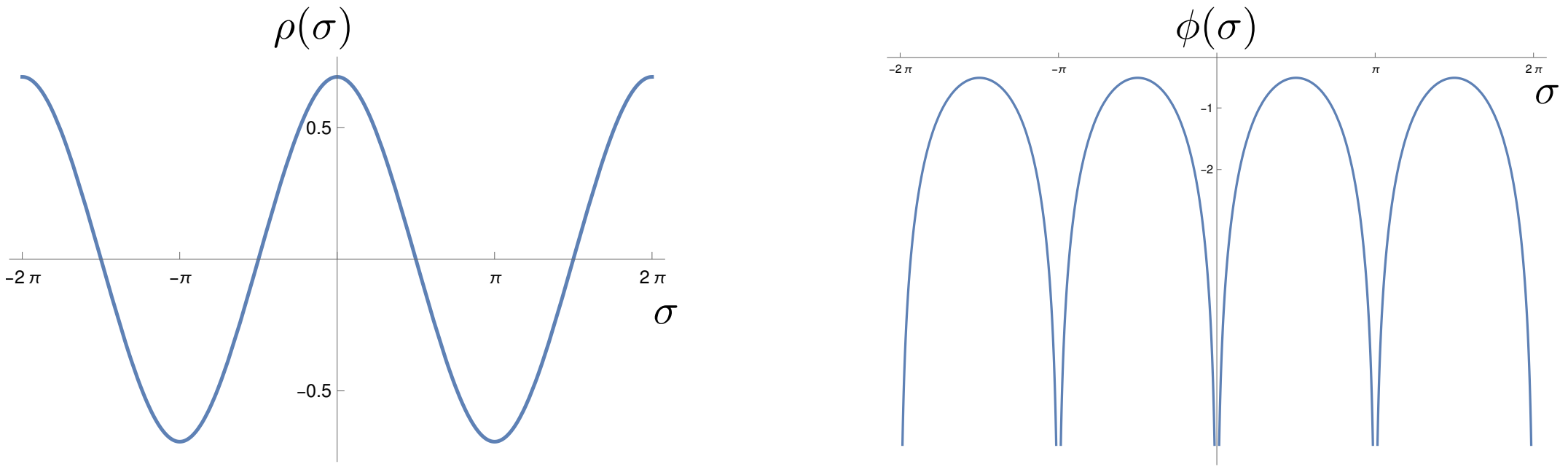
\begin{figure}[ht]
\begin{center}
  \begin{tikzpicture}\node at (0,0){\includegraphics[width=6cm]{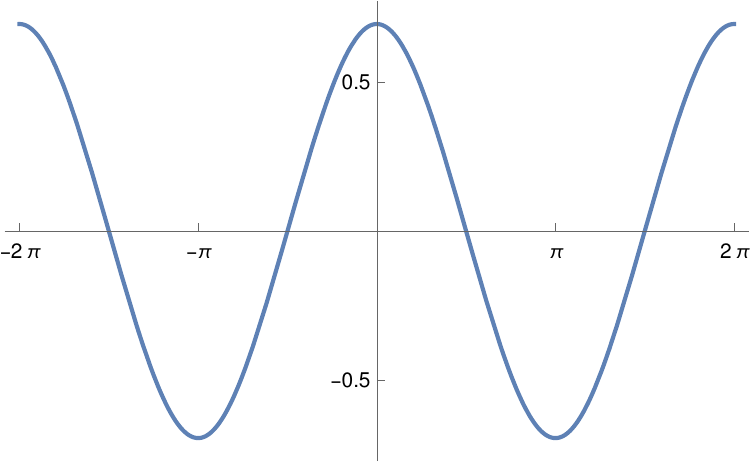}};
    \node at(8,0){\includegraphics[width=6cm]{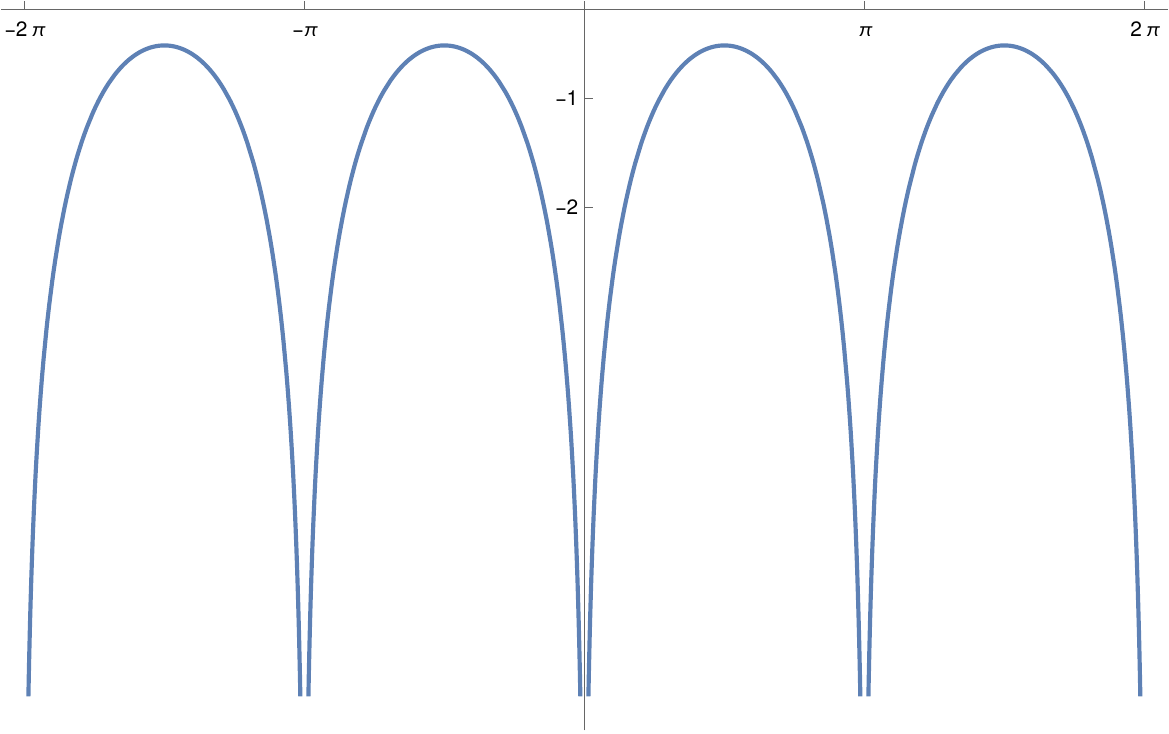}};
\node at (-0.2,2.1){$\rho(\sigma)$};\node at (8,2.1){$\phi(\sigma)$};
    \node at (3,-0.5){$\sigma$};\node at (11,1.5){$\sigma$};
    \node at (0,-2.5){(a)};\node at (8,-2.5){(b)};
  \end{tikzpicture}\end{center}
  \caption{A plot of the (a) the $\AdS_3$ radial coordinate and (b) the Pohlmeyer field for the GKP string. The elliptic parameter has been taken to be $k=0.6$.} \label{Fig:GKP}
\end{figure}

As a check of our computation, let us consider the case of the folded rotating string in $\AdS_3$ \cite{Gubser:2002tv}. In the notation of \cite{Tseytlin:2010jv}, the GKP solution takes the form 
\be \label{GKPansatz}
\rho(\sigma,\tau)=\rho(\sigma)\;,\;t(\sigma,\tau)=\kappa \tau\;,\;\theta(\sigma,\tau)=\omega\tau\;,
\ee
where 
\be\label{GKPsinhr}
\sinh\rho=\frac{k}{\sqrt{1-k^2}}\mathrm{cn}\left(\omega \sigma|k^2\right)\;,
\ee
with $k$ the elliptic parameter, given by $k=\kappa/\omega$. Periodicity of the closed string solution, $\rho(\sigma)=\rho(\sigma+2\pi)$,  requires $\omega=2/\pi K(k^2)$ (and therefore $\kappa=2k/\pi K(k^2)$). With this choice, the function $\rho(\sigma)$ is plotted in Fig. \ref{Fig:GKP} (a). 

As discussed in \cite{Ishii:2023qqd}, the GKP string can also be thought of as an open string solution by cutting it off at an appropriate point on the worldsheet. Given (\ref{GKPansatz}), the $t$ and $\theta$ coordinates are always Neumann. Considering the boundaries to be at $\sigma=-\pi,0$, we see that $\p_\sigma\rho\beq0$, i.e. $\rho$ has a Neumann condition as well.\footnote{In \cite{Ishii:2023qqd} the boundary is at $\sigma=\pm \frac{\pi}{2}$, we have shifted by $\pi/2$ for compatibility with our general discussion above.}  So the GKP string satisfies the NNN boundary condition (as well of course the NDN condition, as $\rho(\sigma)$ is $\tau$-independent). We will focus on the boundary at $\sigma=0$. 

In this simple case one finds that the $u$'s are constant everywhere and equal to:
\be
u^+=-u^-=-\frac{8k K(k^2)^2}{\pi^2}\;,
\ee
while the exponent of the Pohlmeyer field is given by 
\be
e^\phi=\frac{1}{k}\left(1-\mathrm{dn}^2\left(\omega\sigma|k^2\right)\right)\;,
\ee
which vanishes at the boundary $\sigma=0$. It is easy to check that $\phi$ is a solution of the sinh-Gordon equation. For a generic value of $k$, it is plotted in Fig.\ref{Fig:GKP} (b). Like all non-vacuum sinh-Gordon solutions, it is singular. We see that the NNN boundary condition corresponds to the case where $\phi\ra-\infty$ at the boundary, in accordance with the general result above. We also note that $\p_\sigma\phi$ has a discontinuity at $\sigma=0$. 

Although we have found a consistent picture, it could certainly be argued that in the NNN case the integrability of the string sigma model is easier to see directly, rather than by Pohlmeyer reduction, since the Pohlmeyer field necessarily diverges at the boundary.

\subsection{The NND case} \label{Sec:NND}

Let us now move on to impose NND boundary conditions, which correspond to
\be \label{bcNND}
\p_\sigma t(\tau,\sigma)\beq \p_\sigma \rho(\tau,\sigma) \beq 0\;\; ,\;\; \p_\tau\theta(\tau,\sigma)\beq 0\;.
\ee
Of course the Dirichlet conditions also imply that $\p_\tau^2\theta\beq0$ (and similarly for higher $\tau$ derivatives). We also require the preservation of the Neumann conditions, so that e.g. $\p_\tau(\p_\sigma t)\beq0$ and so on.

The NND case was also shown to be integrable in \cite{Ishii:2023ucw} by studying the linear dependence of the PCM currents at the boundary. So, as a check of our methods, we just need to confirm that the boundary conditions on the Pohlmeyer field are consistent with integrability. 

From the equations of motion and Virasoro constraints (and their derivatives at the boundary) one obtains the following restrictions on the coordinates:
\be
\p_\sigma^3t\beq\p_\sigma^3\rho\beq0\;\;,\;\;\p_{\sigma\sigma}\theta\beq0\;.
\ee
The boundary value of $e^\alpha$ can be seen to only depend on $\p_\sigma\theta$:
\be
e^\alpha\beq2\sinh^2\rho (\p_\sigma\theta)^2\;.
\ee
This expression is generically time-dependent, so to determine the boundary condition we need to evaluate $\pt_\sigma\phi$ using (\ref{AdSPohlmeyerbc}).
One finds that $u^+\beq u^-$, and generically the $u$'s don't vanish at the boundary so no limiting procedure is needed. The derivatives of the $u$'s depend on the third derivatives of the Neumann directions, which were determined to vanish at the boundary:
\be
\frac{\p_\sigma u^+}{u^+}+\frac{\p_\sigma u^-}{u^-}\overset{b}{\sim}\left(\p_\tau\rho\p_\sigma^3t-\p_\tau t \p_\sigma^3\rho\right)\beq 0
\ee
We also find
\be
\p_\sigma e^\alpha\beq 2\sinh^2\rho \p_\sigma\theta\p_{\sigma\sigma}\theta\beq 0\;.
\ee
So finally we conclude that:
\be\p_\sigma e^\phi\beq0\;.
\ee
We have obtained a pure Neumann boundary condition for $\phi$, in agreement with the expected integrability of the NND boundary condition \cite{Ishii:2023ucw}.

\subsection{The NDN case} \label{Sec:NDN}

For this case we impose at the boundary
\be
\p_\sigma t(\tau,\sigma)\beq 0\;,\;\p_\tau \rho(\tau,\sigma)\beq0\;,\;\p_\sigma \theta(\tau,\sigma)\beq0\;,
\ee
together with the preservation of the boundary conditions, $\p_\tau(\p_\sigma t)\beq\p_{\tau\tau}\rho\beq\p_{\tau}(\p_{\sigma}\theta)\beq 0$ and so on.

Here the PCM boundary integrability criteria are inconclusive \cite{Ishii:2023ucw}, as there is no linear relation between the boundary currents. However, since (as discussed above) the GKP string is also a special case of the NDN boundary condition, it can be used as a starting solution around which to add fluctuations and track their evolution numerically. This analysis was performed in \cite{Ishii:2023qqd}, with clear signs of turbulence observed. This provides a strong indication that the NDN case is non-integrable. It is therefore natural to wonder whether the corresponding boundary conditions on the Pohlmeyer field are in agreement with this expectation.

The equations of motion imply that at the boundary the Neumann directions satisfy
\be \label{tthetaNDN}
\p_{\tau\tau}t\beq\p_{\sigma\sigma}t\;,\;\p_{\tau\tau}\theta\beq\p_{\sigma\sigma}\theta\;.
\ee
The Dirichlet direction satisfies
\be
\p_{\sigma\sigma}\rho\beq\sinh\rho\cosh\rho((\p_\tau t)^2-(\p_\tau\theta)^2)\;.
\ee
Evaluating (\ref{eA}) we find
\be \label{eaNDN}
e^\alpha\beq\cosh^2\rho(\p_\tau t)^2-\sinh^2\rho(\p_\tau \theta)^2+(\p_\sigma \rho)^2\;,
\ee
while the Virasoro constraints both give
\be \label{virNDN}
-\cosh^2\rho(\p_\tau t)^2+\sinh^2\rho (\p_\tau \theta)^2+(\p_\sigma \rho)^2\beq0\;.
\ee
Combining (\ref{eaNDN}) and (\ref{virNDN}), we obtain
\be
e^\alpha\beq2(\p_\sigma \rho)^2\;.
\ee
This expression is generically nonzero, and time-dependent, at the boundary. Calculating the $u$'s, with the use of (\ref{tthetaNDN}), we obtain
\be\label{uNDN}
\begin{split}
u^++u^-&\beq8e^{-\alpha}\sinh\rho\cosh\rho\left(\p_\tau\theta\p_{\tau\tau}t-\p_\tau t\p_{\tau\tau}\theta\right)\p_\sigma\rho\;,\\
u^+-u^-&\beq-8e^{-\alpha}\p_\tau t\p_\tau\theta(\p_\sigma\rho)^2\;.
\end{split}
\ee
It is easiest to analyse the Pohlmeyer boundary conditions (\ref{AdSPohlmeyerbc}) when there is a definite relation between the $u$'s. At this point it is convenient to restrict our solution space by imposing a simplifying ansatz. This is in the spirit of the analytic non-integrability approach, where finding a single non-integrable example of string motion on a given background is sufficient to rule out integrability of that background. Looking at (\ref{uNDN}), we see that the right-hand-side of $u^+-u^-$ contains precisely the first derivatives left unfixed by the NDN boundary conditions. Ideally, these need to be left generic. Therefore we will look to set $u^++u^-\beq 0$. 

One possible approach is to take the ansatz $\p_\tau t=f(\sigma)\p_\tau \theta$ (not just at the boundary), with $f(\sigma)$ an arbitrary function. This does set $u^++u^-\beq0$, however, it also ends up requiring $e^\phi$ being constant at the boundary. So one is left with an (integrable) Dirichlet boundary condition. Therefore, to look for non-integrability we need to take a different approach.

A hint comes from considering the $\sigma$ derivatives of the equations of motion. Combining the derivatives of (\ref{EomY}) for $\mu=0,1$ and for $\mu=2,3$, and evaluating at the boundary, we find
\be\begin{split} \label{dseomNDN}
&2\sinh\rho\left(\p_{\tau\tau}t\p_\sigma\rho-\p_\tau t \p_{\tau\sigma}\rho\right)+\cosh\rho~ \p_{\sigma}^3t \beq 0\;,\\
&2\cosh\rho\left(\p_{\tau\tau}\theta\p_\sigma\rho-\p_\tau \theta \p_{\tau\sigma}\rho\right)+\sinh\rho ~\p_{\sigma}^3\theta \beq 0\;.
\end{split}\ee
For the Neumann directions  we have $\p_\sigma t \beq \p_\sigma \theta\beq 0$, and from (\ref{tthetaNDN}) their second $\tau$ derivatives are equal to the second $\sigma$ derivatives at the boundary. Therefore, they can be expected to behave like $f(\sigma+\tau)+f(\tau-\sigma)$, with $f$ an even function and where for simplicity the boundary is assumed to be at $\sigma=0$. This guarantees the vanishing of not only the first, but of all odd $\sigma$ derivatives at the boundary. Therefore, as far as the boundary behaviour is concerned, it is natural to impose that $\p_\sigma^3t\beq \p_\sigma^3\theta\beq0$. From (\ref{dseomNDN}) we then obtain:
\be
\p_{\tau\tau}t\beq\frac{\p_{\tau} t\p_{\tau\sigma}\rho}{\p_{\sigma}\rho}\;,\;\p_{\tau\tau}\theta\beq\frac{\p_{\tau} \theta\p_{\tau\sigma}\rho}{\p_{\sigma}\rho}\;.
\ee
Since the second derivatives of $t$ and $\theta$ are proportional to the first ones, from (\ref{uNDN}) we find
\be
u^-\beq-u^+\beq4 e^{-\alpha}\p_\tau t \p_\tau\theta (\p_\sigma\rho)^2\;.
\ee
So, at least close to the boundary, we are in the sinh-Gordon setting. In the following we will take $u^->0$ without loss of generality.
Consistency requires $\p_\tau u^+\beq-\p_\tau u^-$, which can be checked by taking additional time derivatives of the equation of motion. (And applying the boundary conditions consistently, e.g. $\p_{\tau}(\p_{\sigma\sigma}t)\beq\p_\tau (\p_{\tau\tau}t)\beq\p_\tau^3t\beq0,\p_\tau(\p_\sigma^3t)\beq 0$ etc.). Moreover, from the conditions
\be\begin{split}
0&=\p_- u^+=(\p_\sigma u^+-\p_\tau u^+) \;\Rightarrow \;\p_\sigma u^+=\p_\tau u^+ \;,\\
0&=\p_+ u^-=(\p_\sigma u^-+\p_\tau u^-) \;\Rightarrow \;\p_\sigma u^-=-\p_\tau u^-\;,
\end{split}
\ee
we see that at the boundary,
\be\label{ppupmu}
\p_+u^+-\p_-u^- = (\p_\sigma+\p_\tau)u^+-(\p_\sigma-\p_\tau)u^-=2(\p_\tau u^++\p_\tau u^-)\beq 0 \;.
\ee
This can also be verified by directly substituting the equations of motion and their derivatives. 

It is straightforward to compute
\be \label{ephiNDN}
e^\phi=\frac{e^\alpha}{\sqrt{-u^+u^-}}\beq \cosh^2\rho \frac{\p_\tau t}{\p_\tau\theta}-\sinh^2\rho\frac{\p_\tau\theta}{\p_\tau t}\;,
\ee
which is time-dependent as long as $\p_\tau t$ and $\p_\tau\theta$ are not proportional at the boundary.\footnote{As a check of this expression, for the class of ans\"atze (\ref{GKPansatz}) that include the GKP solution,  $\p_\tau \theta=\omega/\kappa \p_\tau t$ and $e^\phi$ becomes constant at the boundary, so all these cases are integrable. For the specific solution (\ref{GKPsinhr}) we have $\omega/\kappa=k$ and $\sinh\rho=\frac{k}{\sqrt{1-k^2}}$ at $\sigma=0$, which implies that $e^\phi=0$, in agreement with the discussion in Section \ref{Sec:NNN}.} 

Moving on to the derivative of $e^\phi$, we calculate
\be\label{dtephiNDN}\begin{split}
\tilde{\p}_\sigma e^\phi&\beq \pt_\sigma \left(\frac{e^\alpha}{\sqrt{-u^+u^-}}\right)\\
&\beq\frac{e^\alpha}{u^-}\pt_\sigma \alpha+\half e^\alpha (\pt_++\pt_-)(-u^+u^-)^{-\half}\\
&\beq e^\phi \pt_\sigma\alpha-\frac{1}{4}\frac{e^\alpha}{\sqrt{-u^+u^-}}\left(\frac{\pt_+u^+}{u^+}+\frac{\pt_-u^-}{u^-}\right)\\
&\beq \frac{e^\phi}{\sqrt{u^-}} \p_\sigma\alpha
+\frac{1}{4}\frac{e^\phi}{(u^-)^{\frac{3}{2}}}\left(\p_+u^+-\p_-u^-\right)\;.
\end{split}
\ee
where in the last line we used that, since $u^+\beq-u^-$, 
\be
\pt_+=\frac{1}{\sqrt{|u^+|}}\p_+\beq\frac{1}{\sqrt{u^-}}\p_+\;,\;\pt_-\beq\frac{1}{\sqrt{u^-}}\p_-\;.
\ee
Given (\ref{ppupmu}), we see that the only contribution to $\pt_\sigma e^\phi$ comes from the first term in (\ref{dtephiNDN}). In terms of the
two independent functions $(\p_\tau t)$ and $(\p_\tau\theta)$, the resulting expression is 
\be \label{psigmaphiNDN}
\begin{split}
\tilde{\p}_\sigma \phi&\beq-\frac{\sinh(2\rho)\left((\p_\tau t)^2-(\p_\tau\theta)^2\right)}{\sqrt{2}\sqrt{(\p_\tau t)(\p_\tau\theta)}
  \sqrt{\cosh^2\rho (\p_\tau t)^2-\sinh^2\rho(\p_\tau\theta)^2}}\\
&\beq -\frac{e^{-\phi}\sinh(2\rho)\left((\p_\tau t)^2-(\p_\tau\theta)^2\right)}{\sqrt{2}(\p_\tau t)(\p_\tau\theta)}\;.
\end{split}
\ee
Since this expression is functionally independent from $e^\phi$ (\ref{ephiNDN}), we can already see that the boundary condition on the Pohlmeyer field is generically not of the Ghoshal-Zamolodchikov form (\ref{GZsinhcosh}). However, it is also instructive to impose a relation between the two independent derivatives, in which case we can invert (\ref{ephiNDN}) and express (\ref{psigmaphiNDN}) in terms of $e^\phi$. Below we do so for two arbitrary relations and expand the derivative in inverse powers of $e^{\phi/2}$ for clarity. 
\be\begin{split}
\p_\sigma\theta\beq (\p_\sigma t)^2 &:\;\; \pt_\sigma\phi\beq -\sqrt2 \tanh\rho e^{\frac{\phi}2}+\frac{\sinh2\rho}{\sqrt2}e^{-\frac{3\phi}2}-\frac{\sinh^32\rho}{4\sqrt{2}} e^{-\frac{7\phi}2}+\cdots\\
\p_\sigma\theta\beq 1/(\p_\sigma t) &:\;\; \pt_\sigma\phi\beq -\sqrt{2}\coth\rho e^{\frac{\phi}2}-\frac{\sinh2\rho}{\sqrt2} e^{-\frac{3\phi}2}+\frac{\sinh^32\rho}{4\sqrt{2}} e^{-\frac{7\phi}{2}}+\cdots
\end{split}
\ee
We note some universality in the subleading terms, which clearly indicates that one can do better and determine this part of the boundary condition without imposing additional relations. However, the non-GZ nature of the boundary condition is already clear, as there is no $e^{-\phi/2}$ term and the series does not truncate. 
  
We have therefore found agreement with the results of \cite{Ishii:2023ucw,Ishii:2023qqd} regarding the non-integrability of the NDN boundary condition. It is noteworthy, however, that the corresponding boundary conditions (i.e. a D-brane positioned at a fixed $\AdS_3$ radial coordinate) were shown to be integrable in Poincar\'e coordinates in \cite{Demjaha:2025axy}. This is not a contradiction, since strings in Poincar\'e coordinates probe a
different region of $\AdS$ and the mapping between the Poincar\'e and global coordinates is non-trivial. Still, it would be important to understand this difference better.

\subsection{The NDD case}

Finally, in this section we consider the case of NDD boundary conditions:
\be \label{bcNDD}
\p_\sigma t(\tau,\sigma)\beq0\;,\; \p_\tau \rho(\tau,\sigma)\beq0\;,\;\p_\tau \theta(\tau,\sigma)\beq0\;,
\ee
where the preservation of the boundary conditions also implies that the $\tau$ derivatives of the above conditions also hold, in particular $\p_{\tau}(\p_\sigma t)\beq0\;,\; \p_{\tau\tau}\rho\beq\p_{\tau\tau}\theta\beq0$.

As discussed in \cite{Ishii:2023ucw}, the PCM integrability criteria are inconclusive in this case. However, a numerical analysis performed in \cite{Ishii:2023ucw} confirmed the presence of turbulence, which was argued to be due to the boundary conditions breaking integrability. It is therefore crucial to confirm whether the Pohlmeyer approach agrees with that conclusion. 

As always, we start with the equations of motion (\ref{EomY}) restricted to the boundary:
\be\begin{split} \label{beomNDD}
\p_{\tau\tau}t&\beq\p_{\sigma\sigma } t\;,\\
\p_{\sigma\sigma}\rho&\beq \sinh\rho\cosh\rho((\p_\tau t)^2+(\p_\sigma \theta)^2)\;,\\
\p_{\sigma\sigma}\theta&\beq-2\coth\rho\p_\sigma\theta\p_\sigma\rho\;.
\end{split}\ee
Imposing the NDD boundary conditions on the Virasoro constraints, we find them to be equal to each other on the boundary and given by:
\be \label{virNDD}
-\cosh^2\rho (\p_\tau t)^2+\sinh^2\rho(\p_\sigma\theta)^2+(\p_\sigma\rho)^2\beq0\;.
\ee
From (\ref{eA}) one finds 
\be \label{eaNDD}
e^\alpha\beq\cosh^2\rho (\p_\tau t)^2+\sinh^2\rho(\p_\sigma\theta)^2+(\p_\sigma\rho)^2\;,
\ee
and combining this with (\ref{virNDD}) we find the following equivalent expressions for $e^\alpha$: 
\be
e^\alpha\beq 2\cosh^2\rho (\p_\tau t)^2\beq 2\left(\sinh^2\rho(\p_\sigma\theta)^2+(\p_\sigma\rho)^2\right)\;.
\ee
As for the $u$'s, a short computation leads to
\be\begin{split}\label{uNDD}
u^++u^-&\beq 4e^{-\alpha}\sinh\rho\cosh\rho (\p_\tau t)^3 \p_\sigma\theta\;,\\
u^+-u^-&\beq -8e^{-\alpha}\sinh\rho\cosh\rho(\p_\tau t)\left(\p_\sigma\rho\p_{\tau\sigma}\theta-\p_\sigma\theta\p_{\tau\sigma}\rho\right)\;.
\end{split}\ee
As in the previous section, in order to proceed it would be convenient to be in the special case where there is a simple relation between the $u$'s. We note that we are in the opposite situation compared to (\ref{uNDN}): The sum $u^++u^-$ depends only on the first derivatives, which we would like to avoid setting to zero. This leaves us with the choice $u^+\beq u^-$. One could try an ansatz of the type $\p_\sigma\rho=f(\sigma) \p_\sigma\theta$, however consistency with the equations of motion means that $f(\sigma)$ has to satisfy an appropriate differential equation. This ends up setting $e^\phi$ to a constant value at the boundary, and therefore leads to a Dirichlet boundary condition. As we are interested in an ansatz general enough to see signs of non-integrability, we need to find an alternative. Let us consider the simple linear ansatz
\be
t(\tau,\sigma)=\kappa\tau\;,
\ee
while $\rho,\theta$ are left arbitrary apart from the boundary conditions (\ref{bcNDD}). Clearly, this ansatz is compatible with the boundary restriction of the $t$ equation of motion (\ref{beomNDD}). It is also compatible with the analysis of \cite{Ishii:2023ucw}, where turbulence was observed within an ansatz where $t$ was kept independent of $\sigma$. 

Substituting $t=\kappa\tau$ in the equations of motion, Virasoro constraints and their derivatives, we find that they imply $\p_{\tau\sigma}\rho\beq \p_{\tau\sigma}\theta\beq0$, which indeed set $u^+=u^-$ in (\ref{uNDD}). One also finds
\be\label{NDDderiv}\begin{split}
&\p_{\sigma\sigma}\theta\beq-2\coth\rho\p_\sigma\theta\p_\sigma\rho\;,\\
&\p_{\sigma\sigma}\rho\beq \sinh\rho\cosh\rho(\kappa^2+(\p_\sigma\theta)^2)\;,\\
&\p_\sigma^3\rho\beq (\kappa^2\cosh(2\rho)-(2+\cosh(2\rho))(\p_\sigma\theta)^2)\p_\sigma\rho\;,\\
&\p_\sigma^3\theta\beq-2(\kappa^2\cosh^2\rho+\cosh^2\rho(\p_\sigma\theta)^2+(1-3\coth^2\rho)(\p_\sigma\rho)^2)\p_\sigma\theta\;.
\end{split}
\ee
The first derivatives are related by (\ref{virNDD}) as $(\p_\sigma\rho)^2+\sinh^2\rho(\p_\sigma\theta)^2=\kappa^2\cosh^2\rho$. We
will use this to rewrite all expressions in terms of $\p_\sigma\rho$. Doing so, one finds that at the boundary
\be \label{eaupmNDD}
e^\alpha\beq2\kappa^2\cosh^2\rho \;\;\text{while}\;\;u^+\beq u^-\beq2\kappa\sinh\rho \sqrt{\kappa^2\cosh^2\rho-(\p_\sigma\rho)^2}\;.
\ee
Therefore, 
\be\label{ephiNDD}
e^\phi\beq \frac{e^\alpha}{u^+}\beq \frac{\kappa\cosh^2\rho}{\sinh\rho\sqrt{\kappa^2\cosh^2\rho-(\p_\sigma\rho)^2}}\;,
\ee
which is generically time-dependent. We then proceed to evaluate the derivative (\ref{AdSPohlmeyerbc})
\be\begin{split} \label{dsEphiNDD}
\pt_\sigma e^\phi&\beq\pt_\sigma \left(\frac{e^\alpha}{\sqrt{u^+u^-}}\right)\\
&\beq e^\phi\pt_\sigma\alpha-\frac{e^\alpha}{4\sqrt{u^+u^-}} \left(\frac{\pt_+u^+}{u^+}+\frac{\pt_-u^-}{u^-}\right)\\
&\beq \frac{e^\phi}{\sqrt{u^+}}\p_\sigma \alpha-\frac{e^\phi}{4(u^+)^{3/2}}(\p_+u^++\p_-u^-)\;.
\end{split}
\ee
where we used (\ref{rescaledderivatives}) on the last line and assumed $u^+>0$ without loss of generality. Now we have:
\be
\p_+u^++\p_-u^-\beq(\p_\sigma+\p_\tau)u^++(\p_\sigma-\p_\tau)u^-\beq \p_\sigma(u^++u^-)+\p_\tau(u^+-u^-)\beq \p_\sigma(u^++u^-)\;,
\ee
since the time derivative of the condition $u^+\beq u^-$ also holds.
It is straightforward to see that 
\be \label{dsEaNDD}
\p_\sigma e^\alpha\beq 4\sinh\rho\cosh\rho\kappa^2\p_\sigma\rho \;\;\Rightarrow\;\; \p_\sigma \alpha\beq 2\tanh\rho ~\p_\sigma\rho\;.
\ee
Using the third derivatives in (\ref{NDDderiv}) as well as the Virasoro constraints, it turns out that
\be
\p_\sigma(u^++u^-)\beq 0\;.
\ee
and therefore $\p_+u^++\p_-u^-=0$.\footnote{Since in this case $\p_\tau u^+\beq\p_\tau u^-$, this result is guaranteed by a variant of the arguments leading up to (\ref{ppupmu}). But it is a good consistency check to also see it explicitly.} 
We conclude that also in the NDD case (with the ansatz $t=\kappa\tau$), the boundary condition comes from the first term in (\ref{dsEphiNDD}).  Combining (\ref{dsEaNDD}) and (\ref{eaupmNDD}), we find
\be
\pt_\sigma \phi\beq\frac{\p_\sigma \alpha}{\sqrt{u^+}}\beq\frac{\sqrt{2}\tanh\rho (\p_\sigma\rho)}{(\kappa \sinh\rho)^{\half}(\kappa^2\cosh^2\rho-(\p_\sigma\rho)^2)^{\frac14}}\;.
\ee
The last step is to invert (\ref{ephiNDD}) in order to express the right-hand-side in terms of $e^\phi$. One finds
\be\begin{split}
\pt_\sigma\phi&\beq\sqrt{2}\tanh\rho e^{\phi/2}\sqrt{1\!-\!\coth^2\rho~e^{-2\phi}}\\
&\beq\sqrt{2}\tanh\rho~e^{\phi/2}\left(1\!-\!\frac{\coth^2\rho}2 e^{-2\phi}\!-\!\frac{\coth^4\rho}6 e^{-4\phi}+\cdots \right)
\end{split}\ee
where in the second line we have expanded in inverse powers of $e^\phi$ for clarity. This boundary condition is not of the Ghoshal-Zamolodchikov form (\ref{GZsinhcosh}). In particular the $e^{-\phi/2}$ term is absent and the series does not truncate. We emphasise that this result is within the $t=\kappa\tau$ ansatz, and therefore we cannot exclude integrable conditions arising if one uses a different ansatz. But the philosophy of our approach (following that of \cite{Basu:2011fw} for closed strings) is that finding a non-integrable example implies that the generic case is non-integrable. We therefore conclude, in accordance with the expectations from \cite{Ishii:2023ucw}, that the NDD boundary condition is generically non-integrable.

\section{Discussion} \label{Section:Discussion}

In this work we studied the boundary conditions imposed on the Pohlmeyer field for prescribed open string boundary conditions, mainly focused on the case of classical strings in $\AdS_3$. We found agreement with the results of \cite{Ishii:2023ucw,Ishii:2023qqd} using alternative approaches (both analytic and numerical). Our main goal was to identify the Pohlmeyer-reduced boundary conditions for the cases where integrability is not obvious, and therefore provide a criterion for \emph{non-integrability} analogous to the methods of \cite{Basu:2011fw} for the closed-string case. 

Although our Pohlmeyer-based approach is not conceptually independent of those based on the PCM (as ultimately both the criteria of \cite{MacKay:2001bh,Delius:2001he} and those of  \cite{Ghoshal:1993tm,MacIntyre:1994sr} boil down to solutions of the boundary reflection equations \cite{Sklyanin:1987bi}) the relative simplicity of the reduced system, and the availability of a complete classification of integrable boundary conditions, might, in some cases, give the Pohlmeyer approach an advantage in deciding the question of classical non-integrability. On the other hand, since the string and reduced systems differ at the quantum level, one needs to be cautious when drawing any conclusions from the Pohlmeyer approach beyond purely classical aspects.\footnote{However, it is conjectured \cite{Roiban:2009vh,Hoare:2009rq} that the original and reduced theories are quantum equivalent in the full $\AdS_5\times\Srm^5$ case where conformal invariance (which is used as input in the reduction) is not broken at the quantum level.} 

Having succeeded in matching with expectations in the simple cases of $Y=0$ non-maximal gravitons and $\AdS_3$, one expects our approach to be applicable to other systems with open strings whose integrability is thus far inconclusive. Understanding the non-maximal $Z=0$ case would be a natural next step. There are several other interesting D-brane configurations in $\AdS_5\times \Srm^5$, such as the case where the D-branes play the role of defects, see e.g. \cite{Okamura:2005cj,Okamura:2007yne, Correa:2008av,Correa:2011nz,Linardopoulos:2021rfq} for discussions of integrable cases and \cite{deLeeuw:2017cop,Linardopoulos:2025ypq} for reviews. From gauge theory considerations, it appears that the specific subspace wrapped by the D-branes can affect the integrability of the configuration. For instance, in the D3-D7 system, the case of the D7-branes wrapping an $\Srm^4$ has been shown to be integrable \cite{DeLeeuw:2018cal}, while that where they wrap an $\Srm^2\times \Srm^2$ appears not to be \cite{deLeeuw:2019sew}. While keeping in mind the above-mentioned caveat regarding extrapolations to the dual theory from a classical equivalence, it would clearly be highly relevant to study the difference between these boundary conditions from the perspective of the Pohlmeyer-reduced system. 

 In our study we have not considered worldvolume fluxes, which are very important for the case of defects above. Turning on fluxes allows for a wider variety of string boundary conditions, which therefore allows for the possibility of different boundary conditions for the reduced theory as well. Giant graviton setups with fluxes have been considered in \cite{Correa:2013em}, where, as mentioned earlier, the Ghoshal-Zamolodchikov condition also appears naturally. Identifying cases where fluxes might lead to a breaking of integrability would be an interesting avenue for future work.
 
Recently, the case of Wilson loops in the Coulomb branch of $\Ncal=4$ SYM was studied in \cite{Demjaha:2025axy}. The setting is equivalent to a D-brane located at a constant distance $z$ from the AdS boundary (located at $z=0$ in Poincar\'e coordinates). That configuration was shown to be integrable, despite the similarity to the NDN case, expected to be non-integrable from the work of \cite{Ishii:2023ucw,Ishii:2023qqd} and our approach as well. The difference is likely due to the nontrivial mapping between Poincar\'e and global coordinates, and it would be interesting to compare the two settings in a more detailed way. 

Making contact with studies of open string motion in more general settings, such as a recent study of chaos on $\AdS$ black hole and $p$-brane backgrounds \cite{Djukic:2023dgk}, would be highly desirable, but appears challenging with our methods, as there is no obvious analogue of Pohlmeyer reduction in those cases. 

In conclusion, we hope that the results presented here provide a useful addition to the available toolkit for the study of integrability and its breaking in the AdS/CFT correspondence.

\paragraph{Acknowledgements} The authors wish to thank Dimitrios Giataganas and George Linardopoulos for insightful discussions. The work of DC and LT was supported by the DSTI-NRF Centre of Excellence in Mathematical and Statistical Sciences (CoE-MaSS). In addition, LT acknowledges a postdoctoral co-funding fellowship from the Department of Research and Innovation (DRI) at the University of Pretoria. Opinions expressed and conclusions arrived at are those of the authors and are not necessarily to be attributed to the CoE-MaSS.

\bibliography{openstrings}

\providecommand{\href}[2]{#2}\begingroup\raggedright\begin{thebibliography}{10}

\bibitem{Maldacena:1997re}
J.~M. Maldacena, ``{The Large N limit of superconformal field theories and
  supergravity},'' \href{http://dx.doi.org/10.1023/A:1026654312961}{{\em Int.
  J. Theor. Phys.} {\bfseries 38} (1999) 1113--1133},
  \href{http://arxiv.org/abs/hep-th/9711200}{{\ttfamily arXiv:hep-th/9711200
  [hep-th]}}.
[Adv. Theor. Math. Phys.2,231(1998)].

\bibitem{Beisert:2010jr}
N.~Beisert {\em et~al.}, ``{Review of AdS/CFT Integrability: An Overview},''
  \href{http://dx.doi.org/10.1007/s11005-011-0529-2}{{\em Lett. Math. Phys.}
  {\bfseries 99} (2012) 3--32},
\href{http://arxiv.org/abs/1012.3982}{{\ttfamily arXiv:1012.3982 [hep-th]}}.

\bibitem{Gubser:2002tv}
S.~S. Gubser, I.~R. Klebanov, and A.~M. Polyakov, ``{A Semiclassical limit of
  the gauge / string correspondence},''
  \href{http://dx.doi.org/10.1016/S0550-3213(02)00373-5}{{\em Nucl. Phys. B}
  {\bfseries 636} (2002) 99--114},
  \href{http://arxiv.org/abs/hep-th/0204051}{{\ttfamily arXiv:hep-th/0204051}}.

\bibitem{Tseytlin:2003ii}
A.~A. Tseytlin, ``{Spinning strings and AdS / CFT duality},'' in {\em {From
  Fields to Strings: Circumnavigating Theoretical Physics: A Conference in
  Tribute to Ian Kogan}}, pp.~1648--1707.
\newblock 11, 2003.
\newblock \href{http://arxiv.org/abs/hep-th/0311139}{{\ttfamily
  arXiv:hep-th/0311139}}.

\bibitem{Plefka:2005bk}
J.~Plefka, ``{Spinning strings and integrable spin chains in the AdS/CFT
  correspondence},'' \href{http://dx.doi.org/10.12942/lrr-2005-9}{{\em Living
  Rev. Rel.} {\bfseries 8} (2005) 9},
\href{http://arxiv.org/abs/hep-th/0507136}{{\ttfamily arXiv:hep-th/0507136
  [hep-th]}}.

\bibitem{Arutyunov:2009ga}
G.~Arutyunov and S.~Frolov, ``{Foundations of the AdS$_5$ $\times$ S$^5$
  Superstring. Part I},''
  \href{http://dx.doi.org/10.1088/1751-8113/42/25/254003}{{\em J. Phys.}
  {\bfseries A42} (2009) 254003},
\href{http://arxiv.org/abs/0901.4937}{{\ttfamily arXiv:0901.4937 [hep-th]}}.

\bibitem{Tseytlin:2010jv}
A.~A. Tseytlin, ``{Review of AdS/CFT Integrability, Chapter II.1: Classical
  AdS5xS5 string solutions},''
  \href{http://dx.doi.org/10.1007/s11005-011-0466-0}{{\em Lett. Math. Phys.}
  {\bfseries 99} (2012) 103--125},
\href{http://arxiv.org/abs/1012.3986}{{\ttfamily arXiv:1012.3986 [hep-th]}}.

\bibitem{Pohlmeyer:1975nb}
K.~Pohlmeyer, ``{Integrable Hamiltonian Systems and Interactions Through
  Quadratic Constraints},'' \href{http://dx.doi.org/10.1007/BF01609119}{{\em
  Commun. Math. Phys.} {\bfseries 46} (1976) 207--221}.

\bibitem{Zoubos:2010kh}
K.~Zoubos, ``{Review of AdS/CFT Integrability, Chapter IV.2: Deformations,
  Orbifolds and Open Boundaries},''
  \href{http://dx.doi.org/10.1007/s11005-011-0515-8}{{\em Lett. Math. Phys.}
  {\bfseries 99} (2012) 375},
\href{http://arxiv.org/abs/1012.3998}{{\ttfamily arXiv:1012.3998 [hep-th]}}.

\bibitem{Mann:2006rh}
N.~Mann and S.~E. Vazquez, ``{Classical Open String Integrability},''
  \href{http://dx.doi.org/10.1088/1126-6708/2007/04/065}{{\em JHEP} {\bfseries
  04} (2007) 065},
\href{http://arxiv.org/abs/hep-th/0612038}{{\ttfamily arXiv:hep-th/0612038
  [hep-th]}}.

\bibitem{Dekel:2011ja}
A.~Dekel and Y.~Oz, ``{Integrability of Green-Schwarz Sigma Models with
  Boundaries},'' \href{http://dx.doi.org/10.1007/JHEP08(2011)004}{{\em JHEP}
  {\bfseries 08} (2011) 004},
\href{http://arxiv.org/abs/1106.3446}{{\ttfamily arXiv:1106.3446 [hep-th]}}.

\bibitem{Linardopoulos:2025ypq}
G.~Linardopoulos, ``{String theory methods for defect CFTs},''
  \href{http://arxiv.org/abs/2501.11985}{{\ttfamily arXiv:2501.11985
  [hep-th]}}.

\bibitem{Sklyanin:1987bi}
E.~K. Sklyanin, ``{Boundary conditions for integrable equations},''
  \href{http://dx.doi.org/10.1007/BF01078038}{{\em Funct. Anal. Appl.}
  {\bfseries 21} (1987) 164--166}.

\bibitem{Sklyanin:1988yz}
E.~K. Sklyanin, ``{Boundary Conditions for Integrable Quantum Systems},''
  \href{http://dx.doi.org/10.1088/0305-4470/21/10/015}{{\em J. Phys. A}
  {\bfseries 21} (1988) 2375--2389}.

\bibitem{Driezen:2021cpd}
S.~Driezen, ``{Modave Lectures on Classical Integrability in 2d Field
  Theories},'' \href{http://dx.doi.org/10.22323/1.404.0002}{{\em PoS}
  {\bfseries Modave2021} (2022) 002},
  \href{http://arxiv.org/abs/2112.14628}{{\ttfamily arXiv:2112.14628
  [hep-th]}}.

\bibitem{MacKay:2001bh}
N.~J. MacKay and B.~J. Short, ``{Boundary scattering, symmetric spaces and the
  principal chiral model on the half line},''
  \href{http://dx.doi.org/10.1007/s00220-002-0735-y}{{\em Commun. Math. Phys.}
  {\bfseries 233} (2003) 313--354},
  \href{http://arxiv.org/abs/hep-th/0104212}{{\ttfamily arXiv:hep-th/0104212}}.
  [Erratum: Commun.Math.Phys. 245, 425--428 (2004)].

\bibitem{Delius:2001he}
G.~W. Delius, N.~J. MacKay, and B.~J. Short, ``{Boundary remnant of Yangian
  symmetry and the structure of rational reflection matrices},''
  \href{http://dx.doi.org/10.1016/S0370-2693(01)01275-8}{{\em Phys. Lett. B}
  {\bfseries 522} (2001) 335--344},
  \href{http://arxiv.org/abs/hep-th/0109115}{{\ttfamily arXiv:hep-th/0109115}}.
  [Erratum: Phys.Lett.B 524, 401--401 (2002)].

\bibitem{Moriconi:2001xz}
M.~Moriconi, ``{Integrable boundary conditions and reflection matrices for the
  O(N) nonlinear sigma model},''
  \href{http://dx.doi.org/10.1016/S0550-3213(01)00527-2}{{\em Nucl. Phys. B}
  {\bfseries 619} (2001) 396--414},
  \href{http://arxiv.org/abs/hep-th/0108039}{{\ttfamily arXiv:hep-th/0108039}}.

\bibitem{Corrigan:1996nt}
E.~Corrigan and Z.-M. Sheng, ``{Classical integrability of the O(N) nonlinear
  sigma model on a half line},''
  \href{http://dx.doi.org/10.1142/S0217751X97001560}{{\em Int. J. Mod. Phys. A}
  {\bfseries 12} (1997) 2825--2834},
  \href{http://arxiv.org/abs/hep-th/9612150}{{\ttfamily arXiv:hep-th/9612150}}.

\bibitem{Aniceto:2017jor}
I.~Aniceto, Z.~Bajnok, T.~Gombor, M.~Kim, and L.~Palla, ``{On integrable
  boundaries in the 2 dimensional $O(N)$ $\sigma$-models},''
  \href{http://dx.doi.org/10.1088/1751-8121/aa8205}{{\em J. Phys. A} {\bfseries
  50} no.~36, (2017) 364002}, \href{http://arxiv.org/abs/1706.05221}{{\ttfamily
  arXiv:1706.05221 [hep-th]}}.

\bibitem{Gombor:2018ppd}
T.~Gombor, ``{New boundary monodromy matrices for classical sigma models},''
  \href{http://dx.doi.org/10.1016/j.nuclphysb.2020.114949}{{\em Nucl. Phys. B}
  {\bfseries 953} (2020) 114949},
  \href{http://arxiv.org/abs/1805.03034}{{\ttfamily arXiv:1805.03034
  [hep-th]}}.

\bibitem{Gil:2026ost}
J.~C. Gil and S.~Driezen, ``{Analytic approach to boundary integrability with
  application to mixed-flux $AdS_3 \times S^3$},''
  \href{http://arxiv.org/abs/2601.09935}{{\ttfamily arXiv:2601.09935
  [hep-th]}}.

\bibitem{Bonomi:2026ygm}
D.~Bonomi and V.~Schomerus, ``{Integrable AdS$_2$ branes in mixed flux
  AdS$_3$},'' \href{http://arxiv.org/abs/2611.10120}{{\ttfamily
  arXiv:2611.10120 [hep-th]}}.

\bibitem{Basu:2011fw}
P.~Basu and L.~A. Pando~Zayas, ``{Analytic Non-integrability in String
  Theory},'' \href{http://dx.doi.org/10.1103/PhysRevD.84.046006}{{\em Phys.
  Rev.} {\bfseries D84} (2011) 046006},
\href{http://arxiv.org/abs/1105.2540}{{\ttfamily arXiv:1105.2540 [hep-th]}}.

\bibitem{MoralesRuiz99}
J.~J. Morales-Ruiz, {\em Differential Galois theory and non-integrability of
  Hamiltonian Systems}.
\newblock Birkhauser, Basel, 1999.

\bibitem{Ghoshal:1993tm}
S.~Ghoshal and A.~B. Zamolodchikov, ``{Boundary S matrix and boundary state in
  two-dimensional integrable quantum field theory},''
  \href{http://dx.doi.org/10.1142/S0217751X94001552}{{\em Int. J. Mod. Phys.}
  {\bfseries A9} (1994) 3841--3886},
  \href{http://arxiv.org/abs/hep-th/9306002}{{\ttfamily arXiv:hep-th/9306002
  [hep-th]}}.
[Erratum: Int. J. Mod. Phys.A9,4353(1994)].

\bibitem{Hofman:2007xp}
D.~M. Hofman and J.~M. Maldacena, ``{Reflecting magnons},''
  \href{http://dx.doi.org/10.1088/1126-6708/2007/11/063}{{\em JHEP} {\bfseries
  11} (2007) 063}, \href{http://arxiv.org/abs/0708.2272}{{\ttfamily
  arXiv:0708.2272 [hep-th]}}.

\bibitem{Ciavarella:2010tp}
A.~Ciavarella, ``{Giant magnons and non-maximal giant gravitons},''
  \href{http://dx.doi.org/10.1007/JHEP01(2011)040}{{\em JHEP} {\bfseries 01}
  (2011) 040}, \href{http://arxiv.org/abs/1011.1440}{{\ttfamily arXiv:1011.1440
  [hep-th]}}.

\bibitem{Correa:2013em}
D.~H. Correa and F.~I. Schaposnik~Massolo, ``{D5-brane boundary reflection
  factors},'' \href{http://dx.doi.org/10.1007/JHEP05(2013)095}{{\em JHEP}
  {\bfseries 05} (2013) 095}, \href{http://arxiv.org/abs/1301.3412}{{\ttfamily
  arXiv:1301.3412 [hep-th]}}.

\bibitem{Ishii:2023ucw}
T.~Ishii, K.~Murata, and K.~Yoshida, ``{Boundary driven turbulence on string
  worldsheet},'' \href{http://dx.doi.org/10.1007/JHEP01(2024)073}{{\em JHEP}
  {\bfseries 01} (2024) 073}, \href{http://arxiv.org/abs/2310.08124}{{\ttfamily
  arXiv:2310.08124 [hep-th]}}.

\bibitem{Ishii:2023qqd}
T.~Ishii, R.~Kitaku, K.~Murata, and C.-M. Yoo, ``{Turbulence on open string
  worldsheets under non-integrable boundary conditions},''
  \href{http://dx.doi.org/10.1007/JHEP02(2024)149}{{\em JHEP} {\bfseries 02}
  (2024) 149}, \href{http://arxiv.org/abs/2310.19317}{{\ttfamily
  arXiv:2310.19317 [hep-th]}}.

\bibitem{Babelon:2003qtg}
O.~Babelon, D.~Bernard, and M.~Talon,
  \href{http://dx.doi.org/10.1017/CBO9780511535024}{{\em {Introduction to
  Classical Integrable Systems}}}.
\newblock Cambridge Monographs on Mathematical Physics. Cambridge University
  Press, 2003.

\bibitem{MacIntyre:1994sr}
A.~MacIntyre, ``{Integrable boundary conditions for classical sine-Gordon
  theory},'' \href{http://dx.doi.org/10.1088/0305-4470/28/4/029}{{\em J. Phys.
  A} {\bfseries 28} (1995) 1089--1100},
  \href{http://arxiv.org/abs/hep-th/9410026}{{\ttfamily arXiv:hep-th/9410026}}.

\bibitem{Baseilhac:2001zd}
P.~Baseilhac and G.~W. Delius, ``{Coupling integrable field theories to
  mechanical systems at the boundary},''
  \href{http://dx.doi.org/10.1088/0305-4470/34/40/304}{{\em J. Phys. A}
  {\bfseries 34} (2001) 8259--8270},
  \href{http://arxiv.org/abs/hep-th/0106275}{{\ttfamily arXiv:hep-th/0106275}}.

\bibitem{Saleur:1994yh}
H.~Saleur, S.~Skorik, and N.~P. Warner, ``{The Boundary sine-Gordon theory:
  Classical and semiclassical analysis},''
  \href{http://dx.doi.org/10.1016/0550-3213(95)00021-J}{{\em Nucl. Phys. B}
  {\bfseries 441} (1995) 421--436},
  \href{http://arxiv.org/abs/hep-th/9408004}{{\ttfamily arXiv:hep-th/9408004}}.

\bibitem{Arthur:2015mva}
R.~Arthur, P.~Dorey, and R.~Parini, ``{Breaking integrability at the boundary:
  the sine-Gordon model with Robin boundary conditions},''
  \href{http://dx.doi.org/10.1088/1751-8113/49/16/165205}{{\em J. Phys. A}
  {\bfseries 49} no.~16, (2016) 165205},
  \href{http://arxiv.org/abs/1509.08448}{{\ttfamily arXiv:1509.08448
  [hep-th]}}.

\bibitem{Corrigan:1994ft}
E.~Corrigan, P.~E. Dorey, R.~H. Rietdijk, and R.~Sasaki, ``{Affine Toda field
  theory on a half line},''
  \href{http://dx.doi.org/10.1016/0370-2693(94)91011-1}{{\em Phys. Lett. B}
  {\bfseries 333} (1994) 83--91},
  \href{http://arxiv.org/abs/hep-th/9404108}{{\ttfamily arXiv:hep-th/9404108}}.

\bibitem{Corrigan:1994np}
E.~Corrigan, P.~E. Dorey, and R.~H. Rietdijk, ``{Aspects of affine Toda field
  theory on a half line},'' \href{http://dx.doi.org/10.1143/PTPS.118.143}{{\em
  Prog. Theor. Phys. Suppl.} {\bfseries 118} (1995) 143--164},
  \href{http://arxiv.org/abs/hep-th/9407148}{{\ttfamily arXiv:hep-th/9407148}}.

\bibitem{Corrigan:1995wx}
E.~Corrigan, ``{Reflections},'' in {\em {7th Regional Conference on Mathemtical
  Physics}}.
\newblock 10, 1995.
\newblock \href{http://arxiv.org/abs/hep-th/9601055}{{\ttfamily
  arXiv:hep-th/9601055}}.

\bibitem{Bowcock:1995vp}
P.~Bowcock, E.~Corrigan, P.~E. Dorey, and R.~H. Rietdijk, ``{Classically
  integrable boundary conditions for affine Toda field theories},''
  \href{http://dx.doi.org/10.1016/0550-3213(95)00153-J}{{\em Nucl. Phys. B}
  {\bfseries 445} (1995) 469--500},
  \href{http://arxiv.org/abs/hep-th/9501098}{{\ttfamily arXiv:hep-th/9501098}}.

\bibitem{McGreevy:2000cw}
J.~McGreevy, L.~Susskind, and N.~Toumbas, ``{Invasion of the giant gravitons
  from Anti-de Sitter space},''
  \href{http://dx.doi.org/10.1088/1126-6708/2000/06/008}{{\em JHEP} {\bfseries
  06} (2000) 008},
\href{http://arxiv.org/abs/hep-th/0003075}{{\ttfamily arXiv:hep-th/0003075
  [hep-th]}}.

\bibitem{Grisaru:2000zn}
M.~T. Grisaru, R.~C. Myers, and O.~Tafjord, ``{SUSY and goliath},''
  \href{http://dx.doi.org/10.1088/1126-6708/2000/08/040}{{\em JHEP} {\bfseries
  08} (2000) 040},
\href{http://arxiv.org/abs/hep-th/0008015}{{\ttfamily arXiv:hep-th/0008015
  [hep-th]}}.

\bibitem{Balasubramanian:2001nh}
V.~Balasubramanian, M.~Berkooz, A.~Naqvi, and M.~J. Strassler, ``{Giant
  gravitons in conformal field theory},''
  \href{http://dx.doi.org/10.1088/1126-6708/2002/04/034}{{\em JHEP} {\bfseries
  04} (2002) 034}, \href{http://arxiv.org/abs/hep-th/0107119}{{\ttfamily
  arXiv:hep-th/0107119}}.

\bibitem{Berenstein:2006qk}
D.~Berenstein, D.~H. Correa, and S.~E. Vazquez, ``{A Study of open strings
  ending on giant gravitons, spin chains and integrability},''
  \href{http://dx.doi.org/10.1088/1126-6708/2006/09/065}{{\em JHEP} {\bfseries
  09} (2006) 065}, \href{http://arxiv.org/abs/hep-th/0604123}{{\ttfamily
  arXiv:hep-th/0604123}}.

\bibitem{Hofman:2006xt}
D.~M. Hofman and J.~M. Maldacena, ``{Giant Magnons},''
  \href{http://dx.doi.org/10.1088/0305-4470/39/41/S17}{{\em J. Phys. A}
  {\bfseries 39} (2006) 13095--13118},
  \href{http://arxiv.org/abs/hep-th/0604135}{{\ttfamily arXiv:hep-th/0604135}}.

\bibitem{Berenstein:2005vf}
D.~Berenstein and S.~E. Vazquez, ``{Integrable open spin chains from giant
  gravitons},'' \href{http://dx.doi.org/10.1088/1126-6708/2005/06/059}{{\em
  JHEP} {\bfseries 06} (2005) 059},
  \href{http://arxiv.org/abs/hep-th/0501078}{{\ttfamily arXiv:hep-th/0501078}}.

\bibitem{Galleas:2009ye}
W.~Galleas, ``{The Bethe Ansatz Equations for Reflecting Magnons},''
  \href{http://dx.doi.org/10.1016/j.nuclphysb.2009.04.024}{{\em Nucl. Phys. B}
  {\bfseries 820} (2009) 664--681},
  \href{http://arxiv.org/abs/0902.1681}{{\ttfamily arXiv:0902.1681 [hep-th]}}.

\bibitem{Nepomechie:2009zi}
R.~I. Nepomechie, ``{Bethe ansatz equations for open spin chains from giant
  gravitons},'' \href{http://dx.doi.org/10.1088/1126-6708/2009/05/100}{{\em
  JHEP} {\bfseries 05} (2009) 100},
  \href{http://arxiv.org/abs/0903.1646}{{\ttfamily arXiv:0903.1646 [hep-th]}}.

\bibitem{MacKay:2010ey}
N.~MacKay and V.~Regelskis, ``{Yangian symmetry of the Y=0 maximal giant
  graviton},'' \href{http://dx.doi.org/10.1007/JHEP12(2010)076}{{\em JHEP}
  {\bfseries 12} (2010) 076}, \href{http://arxiv.org/abs/1010.3761}{{\ttfamily
  arXiv:1010.3761 [hep-th]}}.

\bibitem{Bajnok:2012xc}
Z.~Bajnok, R.~I. Nepomechie, L.~Palla, and R.~Suzuki, ``{Y-system for Y=0 brane
  in planar AdS/CFT},'' \href{http://dx.doi.org/10.1007/JHEP08(2012)149}{{\em
  JHEP} {\bfseries 08} (2012) 149},
  \href{http://arxiv.org/abs/1205.2060}{{\ttfamily arXiv:1205.2060 [hep-th]}}.

\bibitem{Bajnok:2013sza}
Z.~Bajnok, M.~Kim, and L.~Palla, ``{Spectral curve for open strings attached to
  the Y=0 brane},'' \href{http://dx.doi.org/10.1007/JHEP04(2014)035}{{\em JHEP}
  {\bfseries 04} (2014) 035}, \href{http://arxiv.org/abs/1311.7280}{{\ttfamily
  arXiv:1311.7280 [hep-th]}}.

\bibitem{Bak:2011yy}
D.~Bak, B.~Chen, and J.-B. Wu, ``{Holographic Correlation Functions for Open
  Strings and Branes},'' \href{http://dx.doi.org/10.1007/JHEP06(2011)014}{{\em
  JHEP} {\bfseries 06} (2011) 014},
  \href{http://arxiv.org/abs/1103.2024}{{\ttfamily arXiv:1103.2024 [hep-th]}}.

\bibitem{Ciavarella:2010je}
A.~Ciavarella and P.~Bowcock, ``{Boundary Giant Magnons and Giant Gravitons},''
  \href{http://dx.doi.org/10.1007/JHEP09(2010)072}{{\em JHEP} {\bfseries 09}
  (2010) 072}, \href{http://arxiv.org/abs/1007.1674}{{\ttfamily arXiv:1007.1674
  [hep-th]}}.

\bibitem{Ahn:2007bq}
C.~Ahn, D.~Bak, and S.-J. Rey, ``{Reflecting Magnon Bound States},''
  \href{http://dx.doi.org/10.1088/1126-6708/2008/04/050}{{\em JHEP} {\bfseries
  04} (2008) 050}, \href{http://arxiv.org/abs/0712.4144}{{\ttfamily
  arXiv:0712.4144 [hep-th]}}.

\bibitem{Bak:2008xq}
D.~Bak, ``{Zero Modes for the Boundary Giant Magnons},''
  \href{http://dx.doi.org/10.1016/j.physletb.2009.01.035}{{\em Phys. Lett. B}
  {\bfseries 672} (2009) 284--288},
  \href{http://arxiv.org/abs/0812.2645}{{\ttfamily arXiv:0812.2645 [hep-th]}}.

\bibitem{Klose:2008rx}
T.~Klose and T.~McLoughlin, ``{Interacting finite-size magnons},''
  \href{http://dx.doi.org/10.1088/1751-8113/41/28/285401}{{\em J. Phys. A}
  {\bfseries 41} (2008) 285401},
  \href{http://arxiv.org/abs/0803.2324}{{\ttfamily arXiv:0803.2324 [hep-th]}}.

\bibitem{Katsinis:2018zxi}
D.~Katsinis, I.~Mitsoulas, and G.~Pastras, ``{Elliptic string solutions on
  $\mathbb {R}\times \hbox {S}^2$ and their {P}ohlmeyer reduction},''
  \href{http://dx.doi.org/10.1140/epjc/s10052-018-6429-1}{{\em Eur. Phys. J. C}
  {\bfseries 78} no.~11, (2018) 977},
  \href{http://arxiv.org/abs/1805.09301}{{\ttfamily arXiv:1805.09301
  [hep-th]}}.

\bibitem{Katsinis:2018ewd}
D.~Katsinis, I.~Mitsoulas, and G.~Pastras, ``{Dressed elliptic string solutions
  on $\mathbb {R}\times \hbox {S}^2$},''
  \href{http://dx.doi.org/10.1140/epjc/s10052-018-6129-x}{{\em Eur. Phys. J. C}
  {\bfseries 78} no.~8, (2018) 668},
  \href{http://arxiv.org/abs/1806.07730}{{\ttfamily arXiv:1806.07730
  [hep-th]}}.

\bibitem{Katsinis:2019oox}
D.~Katsinis, I.~Mitsoulas, and G.~Pastras, ``{Salient features of dressed
  elliptic string solutions on $\mathbb {R}\times \hbox {S}^2$},''
  \href{http://dx.doi.org/10.1140/epjc/s10052-019-7369-0}{{\em Eur. Phys. J. C}
  {\bfseries 79} no.~10, (2019) 869},
  \href{http://arxiv.org/abs/1903.01408}{{\ttfamily arXiv:1903.01408
  [hep-th]}}.

\bibitem{Chen:2006gea}
H.-Y. Chen, N.~Dorey, and K.~Okamura, ``{Dyonic giant magnons},''
  \href{http://dx.doi.org/10.1088/1126-6708/2006/09/024}{{\em JHEP} {\bfseries
  09} (2006) 024}, \href{http://arxiv.org/abs/hep-th/0605155}{{\ttfamily
  arXiv:hep-th/0605155}}.

\bibitem{Okamura:2006zv}
K.~Okamura and R.~Suzuki, ``{A Perspective on Classical Strings from Complex
  Sine-Gordon Solitons},''
  \href{http://dx.doi.org/10.1103/PhysRevD.75.046001}{{\em Phys. Rev. D}
  {\bfseries 75} (2007) 046001},
  \href{http://arxiv.org/abs/hep-th/0609026}{{\ttfamily arXiv:hep-th/0609026}}.

\bibitem{Grigoriev:2007bu}
M.~Grigoriev and A.~A. Tseytlin, ``{Pohlmeyer reduction of AdS(5) x S**5
  superstring sigma model},''
  \href{http://dx.doi.org/10.1016/j.nuclphysb.2008.01.006}{{\em Nucl. Phys. B}
  {\bfseries 800} (2008) 450--501},
  \href{http://arxiv.org/abs/0711.0155}{{\ttfamily arXiv:0711.0155 [hep-th]}}.

\bibitem{Miramontes:2008wt}
J.~L. Miramontes, ``{Pohlmeyer reduction revisited},''
  \href{http://dx.doi.org/10.1088/1126-6708/2008/10/087}{{\em JHEP} {\bfseries
  10} (2008) 087}, \href{http://arxiv.org/abs/0808.3365}{{\ttfamily
  arXiv:0808.3365 [hep-th]}}.

\bibitem{Hollowood:2009tw}
T.~J. Hollowood and J.~L. Miramontes, ``{Magnons, their Solitonic Avatars and
  the Pohlmeyer Reduction},''
  \href{http://dx.doi.org/10.1088/1126-6708/2009/04/060}{{\em JHEP} {\bfseries
  0904} (2009) 060},
\href{http://arxiv.org/abs/0902.2405}{{\ttfamily arXiv:0902.2405 [hep-th]}}.

\bibitem{Hollowood:2010dt}
T.~J. Hollowood and J.~L. Miramontes, ``{Classical and Quantum Solitons in the
  Symmetric Space Sine-Gordon Theories},''
  \href{http://dx.doi.org/10.1007/JHEP04(2011)119}{{\em JHEP} {\bfseries 04}
  (2011) 119}, \href{http://arxiv.org/abs/1012.0716}{{\ttfamily arXiv:1012.0716
  [hep-th]}}.

\bibitem{Katsinis:2022hyn}
D.~Katsinis, ``{Novel aspects of integrability for nonlinear sigma models in
  symmetric spaces},''
  \href{http://dx.doi.org/10.1103/PhysRevD.105.126008}{{\em Phys. Rev. D}
  {\bfseries 105} no.~12, (2022) 126008},
  \href{http://arxiv.org/abs/2201.10554}{{\ttfamily arXiv:2201.10554
  [hep-th]}}.

\bibitem{Bakas:1995bm}
I.~Bakas, Q.-H. Park, and H.-J. Shin, ``{Lagrangian formulation of symmetric
  space sine-Gordon models},''
  \href{http://dx.doi.org/10.1016/0370-2693(96)00026-3}{{\em Phys. Lett. B}
  {\bfseries 372} (1996) 45--52},
  \href{http://arxiv.org/abs/hep-th/9512030}{{\ttfamily arXiv:hep-th/9512030}}.

\bibitem{Kalousios:2008gz}
C.~Kalousios, G.~Papathanasiou, and A.~Volovich, ``{Exact solutions for
  N-magnon scattering},''
  \href{http://dx.doi.org/10.1088/1126-6708/2008/08/095}{{\em JHEP} {\bfseries
  08} (2008) 095}, \href{http://arxiv.org/abs/0806.2466}{{\ttfamily
  arXiv:0806.2466 [hep-th]}}.

\bibitem{DeVega:1992xc}
H.~J. De~Vega and N.~G. Sanchez, ``{Exact integrability of strings in
  D-Dimensional De Sitter space-time},''
  \href{http://dx.doi.org/10.1103/PhysRevD.47.3394}{{\em Phys. Rev. D}
  {\bfseries 47} (1993) 3394--3405}.

\bibitem{Hayashi:2007bq}
H.~Hayashi, K.~Okamura, R.~Suzuki, and B.~Vicedo, ``{Large Winding Sector of
  AdS/CFT},'' \href{http://dx.doi.org/10.1088/1126-6708/2007/11/033}{{\em JHEP}
  {\bfseries 11} (2007) 033}, \href{http://arxiv.org/abs/0709.4033}{{\ttfamily
  arXiv:0709.4033 [hep-th]}}.

\bibitem{Jevicki:2007aa}
A.~Jevicki, K.~Jin, C.~Kalousios, and A.~Volovich, ``{Generating AdS String
  Solutions},'' \href{http://dx.doi.org/10.1088/1126-6708/2008/03/032}{{\em
  JHEP} {\bfseries 03} (2008) 032},
\href{http://arxiv.org/abs/0712.1193}{{\ttfamily arXiv:0712.1193 [hep-th]}}.

\bibitem{Irrgang:2012xb}
A.~Irrgang and M.~Kruczenski, ``{Rotating Wilson loops and open strings in
  AdS3},'' \href{http://dx.doi.org/10.1088/1751-8113/46/7/075401}{{\em J.
  Phys.} {\bfseries A46} (2013) 075401},
\href{http://arxiv.org/abs/1210.2298}{{\ttfamily arXiv:1210.2298 [hep-th]}}.

\bibitem{Bakas:2016jxp}
I.~Bakas and G.~Pastras, ``{On elliptic string solutions in AdS$_{3}$ and
  dS$_{3}$},'' \href{http://dx.doi.org/10.1007/JHEP07(2016)070}{{\em JHEP}
  {\bfseries 07} (2016) 070}, \href{http://arxiv.org/abs/1605.03920}{{\ttfamily
  arXiv:1605.03920 [hep-th]}}.

\bibitem{Mikhailov:2007xr}
A.~Mikhailov and S.~Schafer-Nameki, ``{Sine-Gordon-like action for the
  Superstring in AdS(5) x S**5},''
  \href{http://dx.doi.org/10.1088/1126-6708/2008/05/075}{{\em JHEP} {\bfseries
  05} (2008) 075}, \href{http://arxiv.org/abs/0711.0195}{{\ttfamily
  arXiv:0711.0195 [hep-th]}}.

\bibitem{Hoare:2009fs}
B.~Hoare and A.~A. Tseytlin, ``{Tree-level S-matrix of Pohlmeyer reduced form
  of AdS(5) x S**5 superstring theory},''
  \href{http://dx.doi.org/10.1007/JHEP02(2010)094}{{\em JHEP} {\bfseries 02}
  (2010) 094}, \href{http://arxiv.org/abs/0912.2958}{{\ttfamily arXiv:0912.2958
  [hep-th]}}.

\bibitem{Hoare:2011fj}
B.~Hoare and A.~A. Tseytlin, ``{Towards the quantum S-matrix of the Pohlmeyer
  reduced version of $AdS_5 \times S^5$ superstring theory},''
  \href{http://dx.doi.org/10.1016/j.nuclphysb.2011.05.016}{{\em Nucl. Phys. B}
  {\bfseries 851} (2011) 161--237},
  \href{http://arxiv.org/abs/1104.2423}{{\ttfamily arXiv:1104.2423 [hep-th]}}.

\bibitem{Hoare:2012nx}
B.~Hoare and A.~A. Tseytlin, ``{Pohlmeyer reduction for superstrings in AdS
  space},'' \href{http://dx.doi.org/10.1088/1751-8113/46/1/015401}{{\em J.
  Phys. A} {\bfseries 46} (2013) 015401},
  \href{http://arxiv.org/abs/1209.2892}{{\ttfamily arXiv:1209.2892 [hep-th]}}.

\bibitem{Demjaha:2025axy}
R.~Demjaha and K.~Zarembo, ``{String integrability on the Coulomb branch},''
  \href{http://dx.doi.org/10.1007/JHEP09(2025)154}{{\em JHEP} {\bfseries 09}
  (2025) 154}, \href{http://arxiv.org/abs/2506.17955}{{\ttfamily
  arXiv:2506.17955 [hep-th]}}.

\bibitem{Roiban:2009vh}
R.~Roiban and A.~A. Tseytlin, ``{UV finiteness of Pohlmeyer-reduced form of the
  AdS(5) x S**5 superstring theory},''
  \href{http://dx.doi.org/10.1088/1126-6708/2009/04/078}{{\em JHEP} {\bfseries
  04} (2009) 078}, \href{http://arxiv.org/abs/0902.2489}{{\ttfamily
  arXiv:0902.2489 [hep-th]}}.

\bibitem{Hoare:2009rq}
B.~Hoare, Y.~Iwashita, and A.~A. Tseytlin, ``{Pohlmeyer-reduced form of string
  theory in AdS(5) x S(5): Semiclassical expansion},''
  \href{http://dx.doi.org/10.1088/1751-8113/42/37/375204}{{\em J. Phys. A}
  {\bfseries 42} (2009) 375204},
  \href{http://arxiv.org/abs/0906.3800}{{\ttfamily arXiv:0906.3800 [hep-th]}}.

\bibitem{Okamura:2005cj}
K.~Okamura, Y.~Takayama, and K.~Yoshida, ``{Open spinning strings and AdS/dCFT
  duality},'' \href{http://dx.doi.org/10.1088/1126-6708/2006/01/112}{{\em JHEP}
  {\bfseries 01} (2006) 112},
\href{http://arxiv.org/abs/hep-th/0511139}{{\ttfamily arXiv:hep-th/0511139
  [hep-th]}}.

\bibitem{Okamura:2007yne}
K.~Okamura, {\em {Aspects of Integrability in AdS/CFT Duality}}.
\newblock PhD thesis, Tokyo U., 2007.
\newblock \href{http://arxiv.org/abs/0803.3999}{{\ttfamily arXiv:0803.3999
  [hep-th]}}.

\bibitem{Correa:2008av}
D.~H. Correa and C.~A.~S. Young, ``{Reflecting magnons from D7 and D5
  branes},'' \href{http://dx.doi.org/10.1088/1751-8113/41/45/455401}{{\em J.
  Phys.} {\bfseries A41} (2008) 455401},
\href{http://arxiv.org/abs/0808.0452}{{\ttfamily arXiv:0808.0452 [hep-th]}}.

\bibitem{Correa:2011nz}
D.~H. Correa, V.~Regelskis, and C.~A.~S. Young, ``{Integrable achiral D5-brane
  reflections and asymptotic Bethe equations},''
  \href{http://dx.doi.org/10.1088/1751-8113/44/32/325403}{{\em J. Phys. A}
  {\bfseries 44} (2011) 325403},
  \href{http://arxiv.org/abs/1105.3707}{{\ttfamily arXiv:1105.3707 [hep-th]}}.

\bibitem{Linardopoulos:2021rfq}
G.~Linardopoulos and K.~Zarembo, ``{String integrability of defect CFT and
  dynamical reflection matrices},''
  \href{http://dx.doi.org/10.1007/JHEP05(2021)203}{{\em JHEP} {\bfseries 05}
  (2021) 203}, \href{http://arxiv.org/abs/2102.12381}{{\ttfamily
  arXiv:2102.12381 [hep-th]}}.

\bibitem{deLeeuw:2017cop}
M.~de~Leeuw, A.~C. Ipsen, C.~Kristjansen, and M.~Wilhelm, ``{Introduction to
  integrability and one-point functions in $\mathcal N=$ 4 supersymmetric
  Yang\textendash{}Mills theory and its defect cousin},''
  \href{http://arxiv.org/abs/1708.02525}{{\ttfamily arXiv:1708.02525
  [hep-th]}}.

\bibitem{DeLeeuw:2018cal}
M.~de~Leeuw, C.~Kristjansen, and G.~Linardopoulos, ``{Scalar one-point
  functions and matrix product states of AdS/dCFT},''
  \href{http://dx.doi.org/10.1016/j.physletb.2018.03.083}{{\em Phys. Lett. B}
  {\bfseries 781} (2018) 238--243},
  \href{http://arxiv.org/abs/1802.01598}{{\ttfamily arXiv:1802.01598
  [hep-th]}}.

\bibitem{deLeeuw:2019sew}
M.~de~Leeuw, C.~Kristjansen, and K.~E. Vardinghus, ``{A non-integrable quench
  from AdS/dCFT},''
  \href{http://dx.doi.org/10.1016/j.physletb.2019.134940}{{\em Phys. Lett. B}
  {\bfseries 798} (2019) 134940},
  \href{http://arxiv.org/abs/1906.10714}{{\ttfamily arXiv:1906.10714
  [hep-th]}}.

\bibitem{Djukic:2023dgk}
V.~Djuki{\'c} and M.~{\v{C}}ubrovi{\'c}, ``{Correlation functions for open
  strings and chaos},'' \href{http://dx.doi.org/10.1007/JHEP04(2024)025}{{\em
  JHEP} {\bfseries 04} (2024) 025},
  \href{http://arxiv.org/abs/2310.15697}{{\ttfamily arXiv:2310.15697
  [hep-th]}}.

\end{thebibliography}\endgroup
\bibliographystyle{utphys}

\end{document}